\documentclass[twocolumn,preprintnumbers,prd,superscriptaddress,nofootinbib,floatfix,showpacs,showkeys]{revtex4-2}

\usepackage{amsmath}
\usepackage{amsfonts}
\usepackage{amssymb}
\usepackage{graphicx}
\usepackage{color}
\usepackage{xcolor}
\usepackage{hyperref}
\usepackage{booktabs}
\usepackage{hyperref}
\usepackage{cleveref}
\usepackage{braket}
\usepackage{mathtools}
\usepackage[rightcaption]{sidecap}
\usepackage{soul}
\usepackage{enumitem}
\usepackage{comment}
\usepackage{physics}
\usepackage{subfigure}
\usepackage{dsfont}
\usepackage{makecell}

\definecolor{vividviolet}{rgb}{0.62, 0.0, 1.0}
\definecolor{amaranth}{rgb}{0.9, 0.17, 0.31}
\definecolor{palatinateblue}{rgb}{0.15, 0.23, 0.89}
\definecolor{brightpink}{rgb}{1.0, 0.0, 0.5}
\definecolor{cornflowerblue}{rgb}{0.39, 0.58, 0.93}
\definecolor{deepcarminepink}{rgb}{0.94, 0.19, 0.22}
\definecolor{radicalred}{rgb}{1.0, 0.21, 0.37}
\hypersetup{ linktoc=all,
    colorlinks, linkcolor={palatinateblue},
    citecolor={brightpink}, urlcolor={amaranth}
}

\newcommand{\be}{\begin{equation}}
\newcommand{\ee}{\end{equation}}
\newcommand{\bs}{\begin{split}}
\newcommand{\bea}{\begin{eqnarray}}
\newcommand{\eea}{\end{eqnarray}}

\newcommand{\bes}{\begin{subequations}}
\newcommand{\ees}{\end{subequations}}

\newcommand{\x}{\mathsf{x}}

\usepackage{tikz,xcolor}
\definecolor{lime}{HTML}{A6CE39}
\DeclareRobustCommand{\orcidicon}{%
	\begin{tikzpicture}
	\draw[lime, fill=lime] (0,0)
	circle [radius=0.16]
	node[white] {{\fontfamily{qag}\selectfont \tiny ID}};
	\draw[white, fill=white] (-0.0625,0.095)
	circle [radius=0.007];
	\end{tikzpicture}
	\hspace{-2mm}
}
\foreach \x in {A, ..., Z}{%
	\expandafter\xdef\csname orcid\x\endcsname{\noexpand\href{https://orcid.org/\csname orcidauthor\x\endcsname}{\noexpand\orcidicon}}
}

\begin{document}
\title{Preserving quantum information  in  $f(Q)$ cosmology}

\author{Salvatore Capozziello\orcidB{}}
\email{capozzie@na.infn.it}
\affiliation{Università degli Studi di Napoli “Federico II”, Dipartimento di Fisica “Ettore Pancini”,
Complesso Universitario di Monte S. Angelo, Via Cinthia Edificio 6, 80126 Napoli, Italy.}
\affiliation{Scuola Superiore Meridionale, Largo San Marcellino 10, 80138 Napoli, Italy.}
\affiliation{
Istituto Nazionale di Fisica Nucleare, Sezione di Napoli,
Complesso Universitario di Monte S. Angelo, Via Cinthia Edificio 6, 80126 Napoli, Italy.}

\author{Alessio Lapponi\orcidA{}}
\email{alessio.lapponi-ssm@unina.it}
\affiliation{Scuola Superiore Meridionale, Largo San Marcellino 10, 80138 Napoli, Italy.}
\affiliation{
Istituto Nazionale di Fisica Nucleare, Sezione di Napoli,
Complesso Universitario di Monte S. Angelo, Via Cinthia Edificio 6, 80126 Napoli, Italy.}

\author{Orlando Luongo\orcidC{}}
\email{orlando.luongo@unicam.it}
\affiliation{Al-Farabi Kazakh National University, Al-Farabi av. 71, 050040 Almaty, Kazakhstan.}
\affiliation{School of Science and Technology, University of Camerino, Via Madonna delle Carceri 9, 62032 Camerino, Italy.}
\affiliation{SUNY Polytechnic Institute, 13502 Utica, New York, USA.}
\affiliation{Istituto Nazionale di Fisica Nucleare (INFN), Sezione di Perugia, Via A.~Pascoli, 06123 Perugia, Italy.}
\affiliation{INAF - Osservatorio Astronomico di Brera, Milano, Italy.}

\author{Stefano Mancini\orcidD{}}
\email{stefano.mancini@unicam.it}
\affiliation{School of Science and Technology, University of Camerino, Via Madonna delle Carceri 9, 62032 Camerino, Italy.}
\affiliation{Istituto Nazionale di Fisica Nucleare (INFN), Sezione di Perugia, Via A.~Pascoli, 06123 Perugia, Italy.}

\begin{abstract}
The effects of  cosmological expansion  on quantum bosonic states are investigated, using quantum information theory. In particular, a generic Bogoliubov transformation of  bosonic field modes is considered and the state change on a single mode is regarded as the effect of a quantum channel. Properties and capacities of this channel are thus explored in the framework of    $f(Q)$ theories. As immediate result, we obtain that the information on a single-mode state appears better preserved, whenever the number of particles produced by the cosmological expansion is small. Hence, similarly to general relativity, we show that analogous particle productions result even if we consider symmetric teleparallel gravity theories. Thus, we investigate a power law $f(Q)$ model, leaving unaltered the effective gravitational coupling, and minimise the corresponding particle production. We thus show how to optimise the preservation of classical and quantum information, stored in a bosonic mode states in the remote past. Finally, we compare our findings with those obtained in general relativity.

\end{abstract}



\maketitle

\section{Introduction}

Data storage is intimately related to the preservation of  information contained in a given set of data \cite{shannon}. A fundamental objective in information storage is ensuring its long-term preservation \cite{Gatenby2007,mancini2019quantum}. To reach this goal, one faces limitations dictated by physical theories \cite{BEKENSTEIN_1990}. Recently, those limitations have stemmed from the natural expansion of the universe \cite{bekenstein2003information,Sidhu_2021}, which can introduce inevitable effects on any physical system as due to the gravitational theory that induces the expansion itself \cite{DOLGOV20031,Faraoni2007}.

To better explain how this works, it is required a comprehensive framework that combines both the underlying gravitational theory and quantum information. This may be provided by relativistic quantum information theory \cite{Mann_2012}, i.e., a treatment that explores a wide range of topics involving the matching of quantum mechanics and general relativity (GR) in communication scenarios, addressing, for example, the problem of causal signaling \cite{Bruckner2014,DeRamon2023}, of black hole information paradox \cite{braunstein2007quantum,chen2022quantum, Capozziello:2024ucm}, of gravity-induced entanglement \cite{van2020quantum,Christodoulou2023} and of  analogue gravity systems \cite{bruschi2013robustness,jacquet2020next}.

The scope of relativistic quantum information  extends to weak gravitational limits, where it examines communication through quantum fields within curved spacetimes \cite{Birrell:1982ix, parker_toms_2009, anastopoulos2023quantum, perche2023role}. Among the most prevalent applications are particle detector models, which represent non-relativistic quantum systems interacting with quantum fields \cite{Unruh1976, Unruh1984, Hu_2012}. These models offer a realistic framework for investigating communication between distant quantum detectors within a diverse range of background spacetimes and trajectories \cite{Brown_2013, Tjoa_2022, Lapponi_2023, lapponi2024making}.

Similarly, one can explore communication in curved spacetimes by tracing the evolution of bosonic or fermionic states from the remote past to the distant future using Bogoliubov transformations \cite{Birrell:1982ix}. This approach is considered both realistic and effective, especially when assessing the preservation of information over extended time intervals \cite{Br_dler_2014, Mancini_2014, Gianfelici_2017, Good_2021}. Further, this communication scenario can be seen as the communication between two particle detector models interacting with the field via the Jaynes-Cummings Hamiltonian \cite{Jaynes1963,Bruschi_2013}.

A specific inquiry, addressing the preservation of information stored in the distant past under the backdrop of cosmological expansion, particularly in the context of fermions, has been recently discussed \cite{Mancini_2014}. In particular, it was proved that the \textit{cosmological particle production} \cite{Ford_2021} is responsible for the damping of a remote past signal. Moreover, the cosmological particle production are the main source of the \textit{cosmological entanglement}, carrying quantum correlations from the remote past \cite{Ball_2006,Fuentes_2010,Mart_n_Mart_nez_2012}. For all these reasons, the particles produced by the expanding universe are expected to have a fundamental role in the storage of classical and quantum information from the remote past, even if encoded in bosonic systems \cite{serafini2017quantum}.

On the other side, the remarkable success of GR has led to insights into our understanding of the universe \cite{2019ACO}. Although the theory is elegant, well passing current observations, there is a theoretical and speculative need for those models extending or modifying Einstein's theory \cite{nojiri2007introduction,Capozziello_2007,De_Felice_2010,capozziello2011extended}. This is crucial for shedding light on potential effects that might manifest in regions of intense gravitational forces where Einstein's theory is thought to break down \cite{lombriser2012cluster,reina2023initial} or in cosmology, where dark energy and dark matter are currently open challenges \cite{nojiri2010future,davari2020testing}.

Among all, a recent theory describing the large scale universe has been employed by virtue of violating the metricity postulate on the metric \cite{ADAK_2013}. Classes of such theories, named  symmetric teleparallel gravity (STEGR), in particular the so called $f(Q)$ gravity \cite{mandal2020cosmography,Jim_nez_2020, Heisenberg:2023lru}, represent attempts entering the so-called \emph{trinity of gravity} \cite{jimenez2019geometrical, Capozziello:2022zzh}, where alternatives to Einstein's theory aim to describe the universe without passing through the concept of curvature. In this respect, promising examples of $f(Q)$ models have been considered in astrophysics \cite{maurya2022anisotropic,bhar2023physical}, gravitation sector \cite{Capozziello:2024vix,Capozziello:2024jir} and cosmology \cite{ALBUQUERQUE2022100980,Khyllep_2023, capozziello2022model,Nojiri:2024zab}, matching several astronomical observations without involving any form of dark energy \cite{Ayuso2021,Anagnostopoulos_2021}.


Motivated by preserving information and developing alternatives to Einstein's gravity, we here investigate preservation of information stored in bosons within the context of $f(Q)$ gravity. In so doing, we study the evolution of bosonic Gaussian states from remote past to far future using Bogoliubov transformation. We show that the rate of classical and quantum information is compromised by the cosmological particle production for bosons. Hence, we seek for the minimization of particle production proving that the latter represents the optimal scenario for  preservation of information stored in remote past. To this aim, we consider modifications of STEGR, wondering if these could better preserve information, stored in bosonic states. Hence, by imposing an appropriate Yukawa-like non-minimal coupling between the phion field and the non-metricity, we find the corresponding rate of particle production. To do so, we specialize our background adopting the Bernard-Duncan scale factor, having the advantage of being flat at past and future regimes. Afterwards, using a precise power law scenario for $f(Q)$, we bound the corresponding free parameters, optimizing the preservation of information from the remote past, indicating how suitable $f(Q)$ theories may be built up. Accordingly, our outcomes  are compared with previous findings, checking possible similarities with both the fermionic case and particles produced in pure GR.


The paper is organized as follows. In Sec.~\ref{sec2}, we study the communication of bosonic Gaussian states under  Bogoliubov transformations, focusing on a single mode. The classical and quantum capacity of the quantum channel arising from these transformations is found as well. In Sec.~\ref{sec3}, we include the universe expansion and see the role of  particles produced on the evolution of bosonic single-mode states. In Sec.~\ref{sec4}, we develop a perturbative approach to study the particle production in STEGR and its $f(Q)$ generalization showing how  gravity modifications could increase or decrease it. A  specific example is provided. Discussion, conclusions and perspectives are reported in Sec.~\ref{conclusioni}\footnote{Throughout the paper, physical units  $c=\hbar=8\pi G=1$ are considered.}.


\section{Evolution of Gaussian states through Bogoliubov transformations}\label{sec2}

Let us examine how a particular class of bosonic states, i.e. Gaussian states, undergoes transformation when subjected to a Bogoliubov transformation. In the context of quantum field theory in curved spacetime, an evolving metric, in the distant past up to the infinite future, involves a Bogoliubov transformation of the normal modes of the bosonic field \cite{Birrell:1982ix,Unruh1976,parker_toms_2009}. As a result, we begin by defining bosonic Gaussian states and exploring their properties. Subsequently, we delve into the transformation of these states under Bogoliubov transformations and we finally examine the properties of a general quantum channel, constructed by the Bogoliubov transformations.

\subsection{Bosonic Gaussian states}\label{ssec2.1}

Among bosonic states, we single out  \textit{bosonic Gaussian states} (BGS), being of paramount importance in \textit{continuous variable quantum information theory} \cite{serafini2017quantum}. We assume possible momenta and the fact that  bosons can  form a discrete set\footnote{The generalization for continuous momenta, in a cosmological expansion context, is straightforward.} of cardinality $N$. We then label each of them with a subscript $i=1\ldots N$.

The expansion of a scalar field $\hat{\Phi}$ into normal modes reads \cite{Birrell:1982ix}
\begin{equation}\label{scalarfield}
\hat{\Phi}=\sum_{i=1}^N\left(a_{i}\phi_i+a_{i}^\dagger\phi_i^\ast\right)\,,
\end{equation}
where $a_\mathbf{i}$ is the annihilation operator associated with a particle with momentum $i$, satisfying the bosonic algebra
\begin{equation}\label{bosalgebra}
    \left[a_i,a^\dagger_{j}\right]=\delta_{ij},\quad\left[a_i,a_j\right]=0\,.
\end{equation}
The set of normal modes $\{\phi_i\}_{i=1}^{N}$ in Eq.~\eqref{scalarfield} has to be complete and orthogonal with respect to the scalar product $\left(\phi_1,\phi_2\right)$ defined as
\begin{align}
(\phi_1,\phi_2)&=-i\int_\Sigma\phi_1(x)\overleftrightarrow{\partial_t}\phi_2^\ast(x)d\Sigma\nonumber\\&\coloneqq-i\int_\Sigma\left(\phi_1(x)\partial_t\phi_2^\ast(x)-(\partial_t\phi_1(x))\phi_2^\ast(x)\right)d\Sigma\,,\label{scproductsc}
\end{align}
where $\Sigma$ is a Cauchy surface for the background spacetime of the field.

An $N$-mode BGS is entirely characterized by two elements:

\begin{itemize}
    \item[1.] the \textit{first momentum vector} $\mathbf{d}=(d_1,\dots,d_N)$, with $d_i=\left(\langle Q_i\rangle,\langle P_i\rangle\right)$,
    \item[2.]  the \textit{covariance matrix}: \begin{equation}\label{completeCovMatrix}
    \sigma=\left(\begin{array}{c|c|c}
       \sigma_{11} & \dots & \sigma_{1N}\\\hline
        \vdots & \ddots & \vdots \\\hline
         \sigma_{N1} & \dots & \sigma_{NN}
    \end{array}\right)\,,
\end{equation}
where\small
\begin{equation}\label{submatrixgeneral}
    \sigma_{ij}=\frac{1}{2}\left(\begin{matrix}
        \langle\left\{Q_i,Q_j\right\}\rangle-2\langle Q_i\rangle\langle Q_j\rangle&        \langle\left\{Q_i,P_j\right\}\rangle-2\langle Q_i\rangle\langle P_j\rangle\\        \langle\left\{P_i,Q_j\right\}\rangle-2\langle P_i\rangle\langle Q_j\rangle&        \langle\left\{P_i,P_j\right\}\rangle-2\langle P_i\rangle\langle P_j\rangle
    \end{matrix}
    \right).
\end{equation}
\end{itemize}

Here,
\begin{align}
Q_i=\frac{1}{\sqrt{2}}\left(a_i^\dagger+a_i\right)\,,
\quad P_i=\frac{1}{i\sqrt{2}}\left(a_i-a_i^\dagger\right)\,,
\end{align}
are called \textit{quadrature operators}, satisfying $\left[Q_i,P_j\right]=i\delta_{ij}$. Moreover, in Eq.~\eqref{submatrixgeneral}, $\langle\cdot\rangle$ indicates the expectation value of the operator $\cdot$ in the bosonic state.

As entropic quantities are independent of the first momentum vector \cite{Adesso_2014}, we set $\mathbf{d}=\mathbf{0}$ for simplicity, having for the submatrices $\sigma_{ij}$ of $\sigma$, described in Eq.~\eqref{submatrixgeneral},
\begin{equation}\label{submatrixgeneralreduced}
    \sigma_{ij}=\frac{1}{2}\left(\begin{matrix}
        \langle\left\{Q_i,Q_j\right\}\rangle&        \langle\left\{Q_i,P_j\right\}\rangle\\        \langle\left\{P_i,Q_j\right\}\rangle&        \langle\left\{P_i,P_j\right\}\rangle
    \end{matrix}
    \right)\,.
\end{equation}
The submatrices $\sigma_{ii}$, in diagonal blocks from the $\sigma$ matrix of Eq.~\eqref{completeCovMatrix}, represent the reduced states of the various modes and, so, from Eq.~\eqref{submatrixgeneralreduced}, expanding the quadrature operators  and using the algebra \eqref{bosalgebra}, each submatrix $\sigma_{ii}$ can be rewritten as
\begin{equation}\label{one-mode cov matrix}
    \sigma_{ii}=\left(\begin{matrix}
        \frac{1}{2}+n_i+\Re m_i&\Im m_i\\\Im m_i&\frac{1}{2}+n_i-\Re m_i
    \end{matrix}\right)\,,
\end{equation}
where $n_i\coloneqq\langle a^\dagger_ia_i\rangle$, i.e., the expectation value of particle number in the mode $i$, and $m_i\coloneqq\langle a_ia_i\rangle$. In this regards, the absolute value of $m_i$ is bounded between $0$ and $\sqrt{n_i(n_i+1)}$ - generating a thermal state and a squeezed vacuum state, respectively.

Conversely, the matrices $\sigma_{i\ne j}$ in the off-diagonal blocks of $\sigma$ represent the correlation between modes $i$ and $j$, so, following the above same procedure, these matrices turn into
\begin{equation}\label{correlation matrix}
    \sigma_{i\ne j}=\left(\begin{matrix}
        \Re \gamma_{ij}+\Re\chi_{ij}&\Im \gamma_{ij}+\Im\chi_{ij}\\-\Im \gamma_{ij}+\Im\chi_{ij}&\Re\gamma_{ij}-\Re \chi_{ij}
    \end{matrix}\right)\,,
\end{equation}
where $\gamma_{ij}=\langle a_i^\dagger a_j\rangle$, $\chi_{ij}=\langle a_ia_j\rangle$.

Then, the parameters defining a $N$-mode Gaussian states become $n_i\in\mathbb{R}$, $m_i,\gamma_{ij},\chi_{ij}\in\mathbb{C}$, with $i,j=1,\dots,N$.

If the modes $i=1,\dots,N$ are not correlated, i.e., $\gamma_{ij}=\chi_{ij}=0$, the covariance matrix $\sigma$, Eq.~\eqref{completeCovMatrix}, leads to $\sigma=\bigoplus_{i=1}^n\sigma_{ii}$. Moreover, a vacuum state is given by $n_i=m_i=\gamma_{ij}=\chi_{ij}=0$, yielding $\sigma=\frac{1}{2}\mathbb{I}_{2N\times 2N}$.

\subsection{Bogoliubov transformation}\label{ssec2.2}

We here investigate the impact of a Bogoliubov transformation on a bosonic Gaussian state as the background spacetime evolves over time, transitioning from remote past to far future, where a static metric is assumed in both of these limiting regions.

In general, the modes at far future (or \textit{output modes}) $\{\phi_i^{out}\}_{i=1}^{N'}$ and the ones at remote past (or \textit{input modes}) $\{\phi_i^{in}\}_{i=1}^N$ are different. Remarkably at the far future, new modes may emerge and, accordingly, the total number of modes, which was originally denoted as $N$, now becomes $N'$. Both sets $\{\phi_i^{in}\}_{i=1}^N$ and $\{\phi_i^{out}\}_{i=1}^{N'}$ consist of complete sets of orthonormal modes.

From this fact, each output mode can be written in terms of the input ones through the completeness relation
\begin{equation}\label{bogtransf}
    \phi_i^{out}=\sum_{j=1}^N\left(\alpha_{ij}\phi_{j}^{in}+\beta_{ij}\phi_{j}^{in\ast}\right)\,,
\end{equation}
dubbed \textit{Bogoliubov transformation}, with  $\alpha_{ij}=\left(\phi_i^{out},\phi_j^{in}\right)$ and $\beta_{ij}=\left(\phi_i^{out},\phi_j^{in\ast}\right)$ named \textit{Bogoliubov coefficients}, defined by the scalar product in Eq.~\eqref{scproductsc}.

The scalar field, at remote past, is expanded in normal modes as in Eq.~\eqref{scalarfield}, replacing $\phi_i$ with $\phi_i^{in}$. Instead, at far future, the field is expanded as
\begin{equation}\label{scalarfieldnew}
\hat{\Phi}=\sum_{i=1}^{N'}\left(b_i\phi_i^{out}+b_i^\dagger\phi_i^{out\ast}\right)\,,
\end{equation}
with new annihilation operator, $b_i$.

Using the equivalence between the right hand sides of Eqs.~\eqref{scalarfield} and \eqref{scalarfieldnew} and expanding the output modes through the Bogoliubov transformation \eqref{bogtransf}, we can find a relation between the input annihilation operator and the output one, say
\begin{equation}\label{invbosbogtransf}
b_i=\sum_{j=1}^N\left(\alpha_{ij}^\ast a_{j}-\beta_{ij}^\ast a_{j}^\dagger\right)\,.
\end{equation}
The relation \eqref{invbosbogtransf} allows to calculate the number of particles on the mode $i$ produced from vacuum, yielding
\begin{equation}\label{particle production general}
    N_i=\bra{0}b_i^\dagger b_i\ket{0}=\sum_{j=1}^N|\beta_{ij}|^2\,,
\end{equation}
where $\ket{0}$ is the remote past vacuum, defined as $a_i\ket{0}=0$. To preserve the  bosonic operators algebra in Eqs. \eqref{bosalgebra}, the Bogoliubov coefficients satisfy the following relations
\begin{subequations}
    \begin{align}
&\sum_{k}^{N}\left(\alpha_{ik}\alpha_{jk}^\ast-\beta_{ik}\beta_{jk}^\ast\right)=\delta_{ij}\,,
\label{bogcond1}\\
&\sum_{k}^N\left(\alpha_{ik}\beta_{jk}-\beta_{ik}\alpha_{jk}\right)=0\,..\label{bogcond2}
    \end{align}
\end{subequations}

If one has a $N$-modes Gaussian state characterized by the parameters $n_i$, $m_i$, $\gamma_{ij}$ and $\chi_{ij}$ and we let the background spacetime evolve in time, the annihilation operators evolve from $a_i$ to $b_i$ according to Eq.~\eqref{invbosbogtransf}. The transformation of the annihilation operators leads to the transformation of the parameters $n_i$, $m_i$, $\gamma_{ij}$ and $\chi_{ij}$ since they are defined no more by $a_i$ but by $b_i$.

In this way, the spacetime evolution leads to a modification of each $N$-modes Gaussian state, modifying the parameters $n_i$, $m_i$, $\gamma_{il}$ and $\chi_{il}$, becoming respectively
\footnotesize
\begin{align}
    &\Tilde{n}_i\coloneqq\langle b_i^\dagger b_i\rangle=\sum_j\left(|\alpha_{ij}|^2n_j-2\Re \left(\alpha_{ij}^\ast\beta_{ij}m_j\right)+|\beta_{ij}|^2(1+n_j)\right)\nonumber\\
    &+\sum_{j\ne k}\left(\alpha_{ij}\alpha_{ik}^\ast\gamma_{jk}-\alpha_{ij}\beta_{ik}^\ast\chi_{jk}^\ast-\beta_{ij}\alpha_{ik}^\ast\chi_{jk}+\beta_{ij}\beta_{ik}^\ast\gamma_{jk}^\ast\right);\label{n tilde}
\end{align}
\begin{align}
    &\Tilde{m}_i\coloneqq\langle b_ib_i\rangle=\sum_j\left(\alpha_{ij}^{\ast 2}m_j-\alpha_{ij}^\ast\beta_{ij}^\ast(1+2n_j)+\beta_{ij}^{\ast 2}m_j^\ast\right)  \nonumber\\&+\sum_{j\ne k}\left(\alpha_{ij}^\ast\alpha_{ik}^\ast\chi_{jk}-\alpha^\ast_{ij}\beta_{ik}^\ast\gamma_{jk}^\ast-\beta_{ij}^\ast\alpha_{ik}^\ast\gamma_{jk}+\beta^\ast_{ij}\beta^\ast_{ik}\chi_{jk}^\ast\right);\label{m tilde}
\end{align}

\begin{align}
    &\Tilde{\gamma}_{il}\coloneqq\langle b_i^\dagger b_l\rangle\nonumber\\&=\sum_j\left(\alpha_{ij}\alpha_{lj}^\ast n_j-\alpha_{ij}\beta_{lj}^\ast m_j^\ast-\beta_{ij}\alpha_{lj}^\ast m_j+\beta_{ij}\beta^{\ast}_{lj}(1+n_j)\right)\nonumber\\
    &+\sum_{j\ne k}\left(\alpha_{ij}\alpha_{lk}^\ast\gamma_{jk}-\alpha_{ij}\beta_{lk}^\ast\chi_{jk}^\ast-\beta_{ij}\alpha_{lk}^\ast\chi_{jk}+\beta_{ij}\beta_{lk}^\ast\gamma_{jk}^\ast\right);\label{gamma tilde}
\end{align}

\begin{align}
    &\Tilde{\chi}_{il}\coloneqq\langle b_ib_l\rangle\nonumber\\&=\sum_j\left(\alpha_{ij}^{\ast}\alpha_{jl}m_j-\alpha_{ij}^\ast\beta_{lj}^\ast(1+n_j)-\beta^\ast_{ij}\alpha_{lj}^\ast n_j+\beta_{ij}^{\ast}\beta_{lj}^\ast m_j^\ast\right)\nonumber\\&+\sum_{j\ne k}\left(\alpha_{ij}^\ast\alpha_{lk}^\ast\chi_{jk}-\alpha^\ast_{ij}\beta_{lk}^\ast\gamma_{jk}^\ast-\beta_{ij}^\ast\alpha_{lk}^\ast\gamma_{jk}+\beta^\ast_{ij}\beta^\ast_{lk}\chi_{jk}^\ast\right)\,.\label{chi tilde}
\end{align}\normalsize

\subsection{One-mode Gaussian channels}\label{ssec2.3}

In this section, focusing on a single mode $i$, we show that the Bogoliubov transformations give rise to a one mode Gaussian quantum channel (OMGC).

A complete classification of OMGCs and the study of their properties or capacities are widely present in the literature \cite{devetak2004capacity,Caruso_2006,Pilyavets_2012,Br_dler_2015}. In this way, the study of the communication properties of the Bogoliubov transformations acting on a single mode would be straightforward.

Let us consider a one-mode Gaussian state, labelling with $i$ the specific mode, represented by the covariance matrix \eqref{one-mode cov matrix}, i.e. $\sigma_{in}$.

Supposing all the other modes in the vacuum state, $n_{j\ne i}=m_{j\ne i}=0$, the covariance matrix after the Bogoliubov transformation $\sigma_{out}$, using Eqs.~\eqref{n tilde} and \eqref{m tilde}, reads
\begin{equation}\label{OMGC input output relation}
    \sigma_{out}=\mathbb{T}^T\sigma_{in}\mathbb{T}+\mathbb{N}\,,
\end{equation}
where
\begin{equation}\label{Tmatrix}
    \mathbb{T}=\left(\begin{matrix}
        \Re\left(\alpha_{ii}-\beta_{ii}\right)&\Im\left(\alpha_{ii}+\beta_{ii}\right)\\- \Im\left(\alpha_{ii}-\beta_{ii}\right)& \Re\left(\alpha_{ii}+\beta_{ii}\right)
    \end{matrix}\right),
\end{equation}
\begin{equation}\label{Nmatrix}
    \mathbb{N}=\left(\begin{matrix}
        \frac{1}{2}\sum_{j\ne i}\left|\alpha_{ij} -\beta_{ij} \right|^2&\sum_{j\ne i}\Im(\alpha_{ij} \beta_{ij} ) \\
        \sum_{j\ne i}\Im(\alpha_{ij} \beta_{ij} ) &\frac{1}{2} \sum_{j\ne i}\left|\alpha_{ij} +\beta_{ij} \right|^2
    \end{matrix}\right).
\end{equation}

Thus, a generic OMGC maps the input $\sigma_{in}$ to the output $\sigma_{out}$ following Eq.~\eqref{OMGC input output relation} with the above matrices, $\mathbb{T}$ and $\mathbb{N}$ \cite{Caruso_2006}.

Moreover, the communication properties of this channel depend only on the determinants of the matrices $\mathbb{T}$ and $\mathbb{N}$. Actually,  $\tau\coloneqq\det\mathbb{T}$ represents the \textit{transmissivity} of the channel, i.e., the amount of input signal transmitted to the output. From Eq.~\eqref{Tmatrix}, $\tau$ reads
\begin{equation}\label{tau}
    \tau=\left|\alpha_{ii}\right|^2-\left|\beta_{ii}\right|^2\,.
\end{equation}
On the other hand, the determinant of the matrix $\mathbb{N}$ is instead related to the additive noise produced by the channel. In particular, calling $\overline{n}$ the average number of noisy particles produced by the channel, we have
\begin{subequations}
    \begin{align}
    \overline{n}&=\sqrt{\det\mathbb{N}}\label{addBnoise}
\,,\qquad\quad {\rm if\,\tau=1,}\\
    \overline{n}&=\frac{\sqrt{\det\mathbb{N}}}{\left|1-\tau\right|}-\frac{1}{2},\quad {\rm \,\,otherwise}\,.\label{additivenoise}
    \end{align}
\end{subequations}

From Eqs.~\eqref{n tilde} and \eqref{m tilde}, we remark that the contributions given by eventual particles present in the remote past, in the modes $j\ne i$, considered as \textit{environment modes}, are independent on $n_i$ and $m_i$. As a consequence, looking at Eq.~\eqref{OMGC input output relation}, it is clear that those contributions would affect the matrix $\mathbb{N}$, i.e., the presence of particles in the environment provides noise to the communication channel.

Further, the communication capabilities of a quantum channel arising from a Bogoliubov transformation and, in particular, the quality of channel communication, is usually quantified by its \textit{capacity}. Accordingly, the \textit{classical capacity (quantum) capacity} of a channel is defined as the maximum rate of classical  (quantum) information that the channel can reliably transmit. The classical capacity for OMGC, under the assumption that Gaussian input states are optimal, reads \cite{Pilyavets_2012}
\begin{align}\label{classical capacity}
    C(E)=&h\left(|\tau|\left(\frac{1}{2}+E\right)+|1-\tau|\left(\frac{1}{2}+\overline{n}\right)\right)\nonumber\\&-h\left(\frac{|\tau|}{2}+|1-\tau|\left(\frac{1}{2}+\overline{n}\right)\right)\,,
\end{align}
where the function $h:\mathbb{R}\to\mathbb{R}$ is defined as
\begin{equation}\label{h function}
    h(x)\coloneqq \left(x+\frac{1}{2}\right)\log_2\left(x+\frac{1}{2}\right)-\left(x-\frac{1}{2}\right)\log_2\left(x-\frac{1}{2}\right)\,.
\end{equation}
To avoid unphysical divergence of the capacity, in Eq.~\eqref{classical capacity}, $E$ has been taken as the upper bound for the number of particles usable to encode the classical message - for this reason, one usually refers to the capacity in Eq.~\eqref{classical capacity} as the \textit{constrained classical capacity}.

Bearing the same assumptions as above, for the quantum capacity, we have \cite{Br_dler_2015}
\begin{equation}\label{quantum capacity}
    Q=\max\left\{0,\,\theta(\tau)\log\left(\frac{\tau}{|1-\tau|}\right)-h\left(\frac{1}{2}+\overline{n}\right)\right\}\,,
\end{equation}
where $\theta$ is the Heaviside step function. Eq.~\eqref{quantum capacity}, differently from Eq.~\eqref{classical capacity}, is not constrained to finite $E$ since it does not diverge.


\section{Communication properties of cosmological expansions}\label{sec3}

To develop the same approach as described above, we here focus on the Bogoliubov transformation arising during the cosmic expansion \cite{BERNARD1977201,Birrell:1982ix,Ford1987,parker_toms_2009,Ford_2021}. To this end, using the procedure introduced in Sec.~\ref{sec2}, we study how an expanding universe can affect information on a single mode state. Hence, specializing to a spatially-flat Friedmann-Lema\^itre-Robertson-Walker  background
\begin{equation}\label{FLRW metric}
    ds^2=-dt^2+a(t)\left(d\mathbf{x}\cdot d\mathbf{x}\right)\,,
\end{equation}
with $a(t)$, the cosmological scale factor.

To guarantee that the Bogoliubov transformations hold, as in Eq.~\eqref{bogtransf}, $a(t)$ might be asymptotically constant at remote past and far future. More generally, if this condition is not fully-satisfied, one could still consider the Bogoliubov transformations by assuming that the vacuum state evolves adiabatically. This may be realized if the cosmological expansion evolves sufficiently slow at infinite past and  future times\footnote{Even though this condition is quite unlikely for realistic cosmological domains, we can consider it as a toy approach to compute the Bogoliubov transformations, see  Ref.~\cite{Ford_2021} for additional details.}.\\

It is convenient to write Eq.~\eqref{FLRW metric} using the confromal time, $\eta$ \cite{Birrell:1982ix}
\begin{equation}\label{conformal FLRW metric}
    ds^2=a^2(\eta)\left(-d\eta^2+d\mathbf{x}\cdot d\mathbf{x}\right)\,,
\end{equation}
where, $dt=a(\eta)d\eta$.

For a scalar field, $\Phi$, enabling a Yukawa-like interaction with curvature, $R$, we may consider the following Lagrangian density
\begin{equation}\label{Lagrangian density GR}
    \mathcal{L}=\frac{1}{2}\left(g^{\mu\nu}\partial_\mu\Phi\partial_\nu\Phi+m^2\Phi^2+\xi R\Phi^2\right)\,,
\end{equation}
where $R=6\left(\frac{\ddot{a}}{a}+\frac{\dot{a}^2}{a^2}\right)$, or
\begin{equation}
    R=6\frac{a''}{a^3}\,,
\end{equation}
where hereafter we denote dots as derivatives w.r.t.~the cosmological time $t$, while primes as derivatives w.r.t.~the conformal time. Finally, the coupling constant, $\xi$, indicates the interaction strength, between the field and the scalar curvature.

From Eq.~\eqref{Lagrangian density GR}, we get the Klein-Gordon equation for one mode with momentum $\mathbf{k}$, i.e.
\begin{equation}\label{KGmassive}
    (\Box +m^2+\xi R)\phi_\mathbf{k}=0\,.
\end{equation}
By using Eq.~\eqref{conformal FLRW metric}, Eq.~\eqref{KGmassive} can be recast as
$\phi_{\mathbf{k}}=\varphi_k(\eta)e^{i\mathbf{k}\cdot\mathbf{x}}$, emphasizing a pure temporal part, $\varphi_k(\eta)$, where $k\coloneqq|\mathbf{k}|$, from a time-independent phase factor $e^{i{\bf k\cdot x}}$.

Because both the modes at remote past and the ones at far future are proportional to $e^{i\mathbf{k}\cdot\mathbf{x}}$, by computing the Bogoliubov coefficients via the scalar product \eqref{scproductsc} one can prove that $\alpha_{\mathbf{k},\mathbf{k}'}\propto\delta_{\mathbf{k},\mathbf{k}'}$ and $\beta_{\mathbf{k}}\propto\delta_{\mathbf{k},-\mathbf{k}'}$. Henceforth, the Bogoliubov coefficients can be written in the form
\begin{subequations}
    \begin{align}
    &\alpha_{\mathbf{k}\mathbf{k}'}=\alpha_k\delta_{\mathbf{k},\mathbf{k}'},\label{alphasimp}\\
    &\beta_{\mathbf{k},\mathbf{k}'}=\beta_k\delta_{\mathbf{k},-\mathbf{k}'}.\label{betasimp}
\end{align}
\end{subequations}
The particles produced from vacuum with momentum $\mathbf{k}$ are then $N_{\mathbf{k}}=N_k=|\beta_k|^2$. So, the condition \eqref{bogcond1} becomes
\begin{equation}\label{condbogdiagonal1}
    |\alpha_{\mathbf{k}}|^2-|\beta_{\mathbf{k}}|^2=1\,.
\end{equation}

\subsection{Information preservation on a single mode}\label{ssec3.1}

We want to study how information about particles from the remote past is preserved during cosmic expansion. Since a cosmological expansion involves a Bogoliubov transformation on the field modes, it plays the role of a one-mode Gaussian channel.

Considering the communication of the single mode, with momentum $\mathbf{k}$, the transmissivity of the channel is given by Eq.~\eqref{tau} using Eqs.~\eqref{alphasimp} and \eqref{betasimp} i.e.
\begin{equation}\label{transmissivity cosmological expansion}
    \tau=|\alpha_k|^2=1+|\beta_k|^2=1+N_k\,,
\end{equation}
where in the second equality we used Eq.~\eqref{condbogdiagonal1}.

Studying the matrix $\mathbb{N}$ from Eq.~\eqref{Nmatrix}, using again Eqs.~\eqref{alphasimp} and \eqref{betasimp}, we have
\begin{equation}
    \det\mathbb{N}=\frac{N_k^2}{4}\,,
\end{equation}
leading to $\overline{n}=0$ from Eq.~\eqref{additivenoise}. Since $\tau>1$ from Eq.~\eqref{transmissivity cosmological expansion}, the cosmological expansion acting on a one-mode bosonic state is a \textit{linear amplifier} channel \cite{holevo1999evaluating}.

The constrained classical capacity and the quantum capacity of the channel can be calculated straightforwardly from Eqs.~\eqref{classical capacity} and \eqref{quantum capacity}, obtaining respectively,

\begin{align}
    C(E)&=h\left(\frac{1}{2}+E+N_k+EN_k\right)-h\left(\frac{1}{2}+N_k\right)\,,\label{constrained classical capacity ce}\\
    \mathcal{Q}&=\log\left(\frac{N_k+1}{N_k}\right)\,.\label{quantum capacity ce}
\end{align}

\begin{figure}[ht]
    \centering
    \includegraphics[scale=0.65]{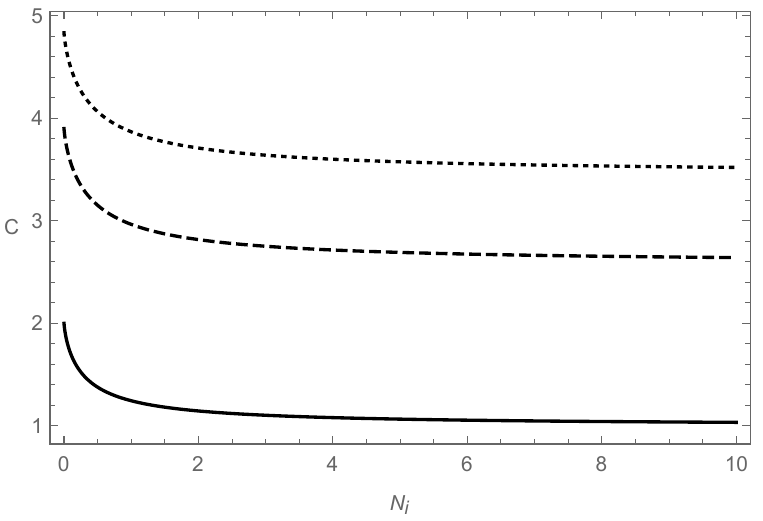}
    \caption{Classical capacity $C$ \eqref{constrained classical capacity ce} of a generic cosmological expansion in terms of the particles produced by it $N_i$ for a single-mode input $i$. The number of encoding particles considered are $E=1$ (thick line), $E=5$ (dashed line), $E=10$ (dotted line).}
    \label{fig:constrained classical capacity one-mode}
\end{figure}
\begin{figure}[ht]
    \centering
    \includegraphics[scale=0.65]{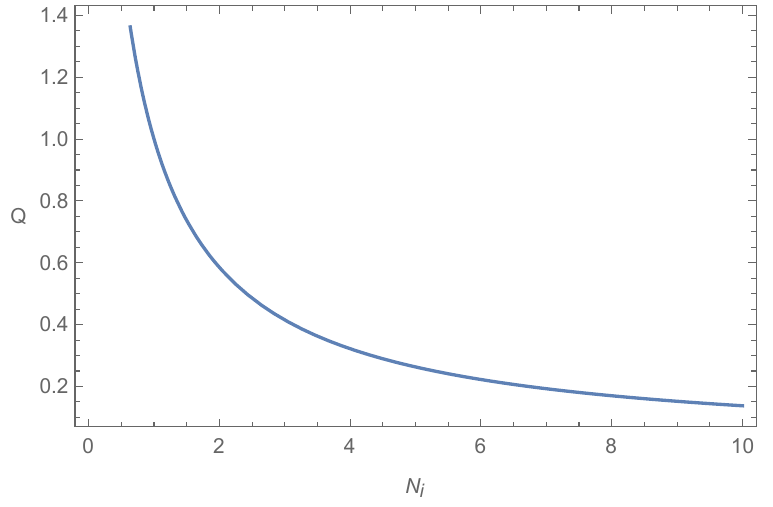}
    \caption{Quantum capacity ${\cal Q}$ \eqref{quantum capacity ce} of a generic cosmological expansion in terms of the particles produced by it $N_i$ for a single-mode input $i$.}
    \label{fig:quantum capacity one-mode}
\end{figure}

The constrained classical capacity, $C(E)$, assumes the value $h\left(\frac{1}{2}+E\right)$ when $N_i=0$ and decreases to $\log(1+E)$ as $N_i\to\infty$. The behaviour of the constrained classical capacity is shown in Fig.~\ref{fig:constrained classical capacity one-mode} for $E=10$, encoding particles. Instead, the quantum capacity is infinite when $N_i=0$ - with no surprise, since in that case we have the identity channel and infinite encoding particles - and drops to zero by increasing the number of particles, as shown in Fig.~\ref{fig:quantum capacity one-mode}.

Concluding, the scenario involving less particle production is the scenario preserving more classical and quantum information from one-mode bosonic Gaussian states from the remote past.

Now, as an interesting comparison, it is worth comparing this result with the case developed in Ref. \cite{Mancini_2014}, where qubit states are communicated instead of Bosonic states. There, a cosmological expansion acting on a qubit leads to a transmissivity $\tau<1$, i.e., an amplitude damping channel. This is due to the different statistics involving bosonic systems and qubit systems. Indeed, reminding that a qubit comes with the occupation numbers in a fermion field momentum mode, the Pauli principle prevents an amplification of the amplitude. Nevertheless, the damping of a qubit with momentum $k$ - and, so, the loss of its classical and quantum information - is proportional to the amount of particles produced in the mode $k$ by the cosmological expansion. Then, in analogy to the bosonic case, also qubit systems' information is better preserved as cosmological particle production is minimized. For the bosonic case, the loss of information is due to an amplification of the input, whereas for the qubit case, it is due to an amplitude damping.

\section{Preserving information in \texorpdfstring{$f(Q)$}{Lg} non-metric gravity cosmology}\label{sec4}

With the above requirements, we wonder whether modified gravity can increase or decrease the preservation of information.

Particularly, as shown in  Ref.~\cite{Capozziello_2016}, the rate of particles produced by a cosmological expansion changes by extending the Hilbert-Einstein action through additional geometric terms.

So, we here focus on STEGR, where curvature is replaced by non-metric term \cite{ADAK_2013,Capozziello:2022zzh}.

Differently from GR, in STEGR, the background is flat and torsionless, and  gravitational effects are due to  non-metricity. In GR, one uses a connection implying a covariant derivative $\nabla_\alpha$ such that $\nabla_\alpha g_{\mu\nu}=0$, i.e. the  \textit{Levi-Civita connection}, while in STEGR, a non-metric connection is considered, say
\begin{equation}\label{non-metricity definition}
    \nabla_{\alpha}g_{\mu\nu}=Q_{\alpha\mu\nu}\,,
\end{equation}
where $Q_{\alpha\mu\nu}$ is the non-metricity tensor. To have STEGR equivalent to GR, the following relation is considered \cite{jimenez2019geometrical, Capozziello:2023vne}
\begin{equation}\label{scalar tensor for RLC}
    R=-Q-\mathcal{D}_\alpha(Q^\alpha-\Tilde{Q}^\alpha)\,,
\end{equation}
where $\mathcal{D_{\alpha}}$ is the covariant derivative w.r.t. the connections and  $Q^\alpha=Q^{\alpha\lambda}_\lambda$, $\Tilde{Q}^\alpha=Q_\lambda^{\lambda\alpha}$, having
\begin{equation}\label{eq: non-metricity scalar}
    Q=\frac{1}{4}Q_{\alpha\beta\gamma}Q^{\alpha\beta\gamma}-\frac{1}{2}Q_{\alpha\beta\gamma}Q^{\beta\alpha\gamma}-\frac{1}{4}Q_\alpha Q^\alpha+\frac{1}{2}Q_\alpha \Tilde{Q}^\alpha\,,
\end{equation}
providing the Hilbert-Einstein action to be
\begin{equation}\label{einstein hilbert action STEGR}
    S_{STEGR}=-\frac{1}{2}\int \sqrt{-g}Qd^4x\,.
\end{equation}

\subsection{Coupling with  non-metricity}\label{ssec4.1}

Following the same recipe developed to arrive to Eq.~\eqref{Lagrangian density GR}, with the aim of investigating cosmological particle production in STEGR, it is convenient to introduce a coupling between the scalar field and the non-metricity tensor $Q_{\alpha\mu\nu}$ defined in Eq.~\eqref{non-metricity definition}. This technique permits to understand which effects a non-minimal coupling between non-metricity and fields can provide.

To this end, we remind that the covariant derivative $\nabla_\alpha$ defining $Q_{\alpha\mu\nu}$ from Eq.~\eqref{non-metricity definition} is not unique, leading to a gauge choice. Among all, we choose the so-called \textit{coincident gauge}, where the connection vanishes globally and where the metricity tensor is expressed via partial derivatives \cite{Jim_nez_2018,Bahamonde_2022} through\small
\begin{equation}
\frac{1}{2}g^{\alpha\lambda}\left(g_{\lambda\nu,\mu}+g_{\mu\lambda,\nu}-g_{\mu\nu,\lambda}\right)=\frac{1}{2}g^{\alpha\lambda}\left(Q_{\mu\alpha\nu}+Q_{\nu\alpha\mu}-Q_{\alpha\mu\nu}\right)\,.\label{coincident gauge relation}
\end{equation}\normalsize
In this case, the non-metricity scalar, $Q$, reads \cite{Jim_nez_2020,ALBUQUERQUE2022100980,Khyllep_2023}
\begin{equation}\label{non-metricity scalar FLRW metric}
    Q=-g^{\mu\nu}\left(\left\{{}_{\beta\mu}^{\alpha}\right\}\left\{{}_{\nu\alpha}^\beta\right\}-\left\{{}^\alpha_{\beta\alpha}\right\}\left\{{}^\beta_{\mu\nu}\right\}\right)\,.
\end{equation}\normalsize
This choice is particularly convenient when studying cosmological expansions. Indeed, considering Eq.~\eqref{FLRW metric} and using Eq.~\eqref{non-metricity scalar FLRW metric}, one infers
\begin{equation}
    Q=6\left(\frac{\dot{a}}{a}\right)^2=6H^2\,,
\end{equation}
where $H$ is the \textit{cosmological Hubble parameter}.

Hence, similarly to Eq.~\eqref{Lagrangian density GR}, we can assume the following interaction
\begin{equation}\label{ansatz lagrangian STEGR}
    \mathcal{L}=\frac{1}{2}\left(g^{\mu\nu}\partial_\mu \Phi\partial_\nu\Phi+m^2\Phi^2+\zeta Q\Phi^2\right)\,,
\end{equation}
where $\zeta$ is a dimensionless coupling constant, indicating the strength of the Yukawa-like interaction between $Q$ and $\Phi$. The modified Klein-Gordon equation reads
\begin{equation}
    (\Box -m^2)\Phi-\zeta Q\Phi=0\,.
\end{equation}
Expanding the field into normal modes, as in Eq.~\eqref{scalarfield}, one gets
\begin{equation}\label{KGmassive with Q}
    \Box \phi_\mathbf{k}-m^2\phi_\mathbf{k}-\zeta Q\phi_\mathbf{k}=0\,,
\end{equation}
that can be recast using the conformal time,  $\eta$, defined in Eq.~\eqref{conformal FLRW metric}. Precisely, the Hubble parameter is ${\displaystyle H=\frac{a'}{a^2}}$ and
\begin{equation}\label{Hubble derivative}
    H'=\frac{a''}{a^2}-2\frac{a'^2}{a^3}\,,
\end{equation}
so, the d'Alemebert operator $\Box$, defined with the covariant derivatives of the Levi-Civita connection w.r.t. the metric \eqref{conformal FLRW metric}, becomes
\begin{equation}
    \Box=-\frac{1}{a^4}\left(2a'a\partial_\eta+a^2\partial^2_\eta\right)+\nabla^2\,.
\end{equation}
The wavefunction $\phi_\mathbf{k}(\mathbf{x},\eta)$ can be rewritten in the following way
\begin{equation}\label{wave function decomposition}
    \phi_{\mathbf{k}}(\mathbf{x},\eta)=\frac{e^{i\mathbf{k}\cdot\mathbf{x}}}{a(\eta)\sqrt{(2\pi)^3}}\chi_k(\eta)\,.
\end{equation}
Recalling,
\begin{subequations}
    \begin{align}
    &\left(\frac{\chi_k}{a}\right)'=\frac{\chi_k' a-a'\chi_k}{a^2}\,,\\
    &    \left(\frac{\chi_k}{a}\right)''=\frac{a^3\chi_k''-a^2a''\chi_k-2a^2a'\chi'_k+2aa'^2\chi_k}{a^4}\,,
    \end{align}
\end{subequations}
from Eq.~ \eqref{KGmassive with Q}, we obtain
\begin{equation}\label{Eq. pre KG}
    \chi''_k+k^2\chi_k+a^2\left(m^2-\frac{a''}{a^3}+6\zeta\frac{a'^2}{a^4}\right)\chi_k=0\,.
\end{equation}
Then, by means of Eq.~\eqref{Hubble derivative}, we can write
\begin{equation}
    \frac{a''}{a^3}=\frac{H'}{a}+2H^2\,,
\end{equation}
so that Eq.~\eqref{Eq. pre KG} becomes
\begin{equation}\label{final KG momentum}
\chi''_k+k^2\chi_k+a^2\left(m^2+2(3\zeta-1)H^2-\frac{H'}{a}\right)=0\,.
\end{equation}
Eq.~\eqref{final KG momentum} can be rewritten using $\omega_{in}^2=k^2+a_{in}^2m^2$, with $a_{in}\coloneqq a(\eta\to-\infty)$, namely
\begin{equation}\label{final KG frequency}
    \chi''_k(\eta)+\omega_{in}^2\chi_k(\eta)+U(\eta)\chi_k(\eta)=0\,,
\end{equation}
where
\begin{equation}\label{potential always valid}
    U(\eta)\equiv m^2(a^2(\eta)-a_{in}^2)+a^2\left(2(3\zeta-1)H^2(\eta)-\frac{H'(\eta)}{a(\eta)}\right)\,.
\end{equation}

\subsection{Perturbative particle production}\label{ssec4.2}

It is clear that Eq.~\eqref{final KG frequency} cannot be solved analytically. Consequently, we here present a possible strategy to perturbatively compute it, following  Refs.~\cite{Zeldovich:1971mw,Birrell:1982ix,Ford_2021}. Despite this method was applied as $\Phi$ couples with the scalar curvature, $R$, the subsequent generalization where $Q$ replaces $R$ appears straightforward, in the coincident gauge.

Hence, for $\eta\to-\infty$  the Minkowski spacetime is reached, so that $\chi_k(\eta)\to \frac{e^{-i\omega_{in}\eta}}{\sqrt{2\omega_{in}}}$ and, using this recipe  as a boundary condition, Eq.~ \eqref{final KG frequency} can be rewritten as
\begin{equation}\label{integral version of KG equation}
    \chi_k(\eta)=\frac{e^{-i\omega_{in}\eta}}{\sqrt{2\omega_{in}}}-\int_{-\infty}^\eta U(\eta_1)\frac{\sin(\omega_{in}(\eta-\eta_1))}{\omega_{in}}\chi_k(\eta_1)d\eta_1\,.
\end{equation}
This relation can be solved by iteratively substituting $\chi_k(\eta_1)$ inside the right hand side of Eq.~\eqref{integral version of KG equation}.

From now on, we assume that $|U(\eta)|\ll\omega_{in}^2$ for each $\eta$. We will verify this assumption \emph{a posteriori} in the text. In such a way, we can neglect terms involving a product of many $U(\eta)/\omega_{in}^2$, having a perturbative solution for $\chi_k(\eta)$. Hence, the zeroth, first and second order read\small
\begin{subequations}
\begin{align}
    &\chi_k^{(0)}(\eta)=\frac{e^{-i\omega_{in}\eta}}{\sqrt{2\omega_{in}}}\,,\\
    &\chi_k^{(1)}(\eta)=\frac{i}{(2\omega_{in})^{3/2}}\int_{-\infty}^\eta\left(e^{i\omega_{in}(\eta-2\eta_1)}-e^{-i\omega_{in}\eta}\right)U(\eta_1)d\eta_1\,,\\
    &\chi_k^{(2)}(\eta)=-\frac{1}{\sqrt{2\omega_{in}^5}}\int_{-\infty}^{\eta}d\eta_1\sin(\eta-\eta_1)U(\eta_1)\times\\
    &\int_{-\infty}^{\eta_1}d\eta_2\sin(\eta_1-\eta_2)U(\eta_2)e^{-i\omega_{in}\eta_2}\,.\nonumber
\end{align}
\end{subequations}\normalsize

Up to the second order, the far future solution for $\chi_k(\eta)$ is then,
\begin{equation}\label{far future chi}
    \chi_{k}(\eta\to+\infty)=e^{-i\omega_{in}\eta}+\chi_k^{(1)}(\eta\to+\infty)+\chi_k^{(2)}(\eta\to+\infty)\,.
\end{equation}

The output modes are thus obtained by inserting Eq.~\eqref{far future chi} into Eq.~\eqref{wave function decomposition}, obtaining
{\small
\begin{align}
    \phi_{\mathbf{k}}^{out}(\mathbf{x},\eta)=\frac{e^{i\mathbf{k}\cdot\mathbf{x}}}{a_{f}\sqrt{(2\pi)^3}}&\left(e^{-i\omega_{in}\eta}+\chi_k^{(1)}(\eta\to+\infty)\right.\nonumber\\&\left.+\chi_k^{(2)}(\eta\to+\infty)\right)\,,
\end{align}
}
where $a_f\coloneqq a(\eta\to\infty)$, whereas the input modes are
\begin{equation}
    \phi_{\mathbf{k}}^{in}(\mathbf{x},\eta)=\frac{e^{i\mathbf{k}\cdot\mathbf{x}}}{a_{in}\sqrt{(2\pi)^3}}\,.
\end{equation}
Using now Eq.~\eqref{scproductsc}, the Bogoliubov coefficients, Eqs.~\eqref{alphasimp} - \eqref{betasimp}, can be found as
\begin{widetext}
\begin{subequations}
    \begin{align}
    &\alpha_k=\frac{a_f}{a_{in}}\left(1+\frac{i}{2\omega_{in}}\int_{-\infty}^{+\infty}d\eta_1U(\eta_1)+\frac{1}{4\omega_{in}^2}\int_{-\infty}^{+\infty}\int_{-\infty}^{+\infty}d\eta_1d\eta_2U(\eta_1)U(\eta_2)(e^{2\omega_{in}i(\eta_1-\eta_2)}-1)\right)\,,\\
    &\beta_k=-\frac{a_f}{a_{in}}\left(\frac{i}{2\omega_{in}}\int_{-\infty}^{+\infty}d\eta_1e^{-2i\omega_{in}\eta_1}U(\eta_1)+\frac{1}{4\omega_{in}^2}\int_{-\infty}^{+\infty}\int_{-\infty}^{+\infty}d\eta_1 d\eta_2e^{2\omega_{in}i(\eta_2-\eta_1)}U(\eta_1)U(\eta_2)\right)\,.
\end{align}
\end{subequations}
\end{widetext}
The number of produced particles then reads
\begin{equation}\label{particle production perturbative}
    N_k=|\beta_k|^2=\frac{a_f^2}{4a_{in}^2\omega_{in}^2}\left|\int_{-\infty}^{+\infty}d\eta_1e^{-2i\omega_{in}\eta_1}U(\eta_1)\right|^2\,.
\end{equation}

\subsection{Affecting particle production through $f(Q)$}\label{ssec4.3}

The STEGR extensions involve analytical functions of $Q$, say $f(Q)$, in lieu of $Q$, modifying the action,  Eq.~\eqref{einstein hilbert action STEGR}, as \cite{mandal2020cosmography,Jim_nez_2020}
\begin{equation}
     S_{f(Q)}=-\frac{1}{2}\int \sqrt{-g}f(Q)d^4x\,.
\end{equation}
In this case, the Friedmann equations are \cite{jimenez2019geometrical,ALBUQUERQUE2022100980,capozziello2022model}
\begin{subequations}\label{friedmann equations modified gravity}
    \begin{align}
        &H^2=\frac{\rho+\frac{1}{2}f(Q)}{6f_Q}\,,\\
        &\frac{H'}{a}=-\frac{1}{2}\frac{\rho+p}{12H^2f_{QQ}+f_Q}\,,
    \end{align}
\end{subequations}
where $\rho$ and $p$ are the density and pressure, respectively. Clearly, we can recover the standard Friedmann equations in GR, as $f(Q)\rightarrow Q=6H^2$.

From Eq.~\eqref{friedmann equations modified gravity}, the inclusion of a further $f(Q)$ fluid implies a different evolution of the Hubble parameter, $H(\eta)$, and, then, from Eqs.~ \eqref{particle production perturbative} and \eqref{potential always valid}, an expected different particle production. Namely, fixing the parameters $\rho$ and $p$, Eq.~\eqref{friedmann equations modified gravity} gives the new solution for the Hubble parameter in terms of $\eta$. Since $H(\eta)$ changes, also the scale factor $a(\eta)$ changes depending on the $f(Q)$ theory chosen, becoming
\begin{equation}\label{modified scale factor}
    a_{mod}(\eta)=\left(\frac{1}{a_{in}}-\frac{1}{\sqrt{6}}\int_{-\infty}^{\eta}\sqrt{\frac{\rho(\eta')+\frac{1}{2}f(Q)}{f_Q}}d\eta'\right)^{-1}\,.
\end{equation}\normalsize
In turn, the function $U(\eta)$ from Eq.~\eqref{potential always valid} becomes
\begin{widetext}
    \begin{equation}\label{potential modified gravity}
        U(\eta)\coloneqq m^2(a_{mod}^2(\eta)-a_{in}^2)+a^2_{mod}(\eta)\left((3\zeta-1)\frac{\rho+\frac{1}{2}f(Q)}{3f_Q}+\frac{f_Q}{2}\frac{\rho+p}{2\rho f_{QQ}+f(Q)f_{QQ}+f_Q^2}\right)\,.
    \end{equation}
\end{widetext}

The particle production in a modified gravity theory $f(Q)$ can be computed plugging $U(\eta)$ from Eq.~\eqref{potential modified gravity} into Eq.~\eqref{particle production perturbative}. Moreover, in Eq.~\eqref{particle production perturbative}, $a_f$ is now given by $a_{mod}(\eta\to+\infty)$.

Among the modified gravity theories $f(Q)$, only a part of them involves a finite particle production from remote past to future infinity. Indeed, from Eq.~\eqref{particle production perturbative}, to have a finite result for $N_k$, we need $a_{mod}$, from Eq.~\eqref{modified scale factor}, to be asymptotically constant for $\eta\to0$. This implies that $H^2$ drops to zero at infinite future and, so, since $Q=6H^2$, we need   $f(Q)$ well-behaves as $Q\to0$.

Moreover, from local observations of the gravitational constant, recalling that in $f(Q)$ theories\footnote{The equality adopts the choice $8\pi G=1$.}, we have \cite{jimenez2019geometrical,Anagnostopoulos_2021}
\begin{equation}\label{effective gravitational coupling}
    8\pi G_{eff}\equiv\frac{1}{f_Q(Q)}\,.
\end{equation}
To require consistency in the Solar System, we need $8\pi G_{eff}$ not to deviate from unity, at far future, i.e. for $Q\to0$.

Then, to allow $f(Q)$ theories to have a finite particle production, we need to select suitable models. For example, we may consider a power-law approach, \cite{Jim_nez_2020,Ayuso2021,Khyllep_2023},
\begin{equation}\label{power-law modification}
    f(Q)=Q+\epsilon A \left(\frac{Q}{A}\right)^d\,,
\end{equation}
where $A>0$. The above scenario reduces to the standard $\Lambda$CDM model for $d=0$ and appears quite consistent with the constraints imposed from Big Bang Nucleosynthesis \cite{Anagnostopoulos_2023,Benetti:2020hxp, Capozziello:2017bxm}. In the far future, $f(Q)$ is finite if $d\ge0$ and for $8\pi G_{eff}$ we have
\begin{equation}
    8\pi G_{eff}=\begin{cases}
        &\frac{1}{1+\epsilon}\quad\text{if}\quad d=1\,;\\
        &1 \quad \text{if}\quad d=0 \quad\text{or}\quad d>1\,;\\
        &0\quad \text{otherwise}\,.
    \end{cases}
\end{equation}
Then, the parameters in Eq.~\eqref{power-law modification} can be either $d=0$ or $d\ge1$, where the equality holds as long as $\epsilon\ll1$.

\section{Particle production in asymptotically flat spacetimes}\label{ssec:4.4}

Besides the choices of $f(Q)$, particle production requires spacetimes to be flat at asymptotic regimes. In Einstein-de Sitter universes, this cannot be easily accounted, i.e., as the universe is dominated by precise fluids, there is no chance to guarantee the flatness. To this end, we may now consider a very precise example of cosmological expansion in which particle production easily occurs.

The model has been firstly presented in Ref.~\cite{BERNARD1977201} through the proposed scale factor,
\begin{equation}\label{BD model}
    a(\eta)=\frac{1}{\sqrt{2}}\sqrt{\left(1+a_{in}^2+(1-a_{in}^2)\tanh(r\eta)\right)}\,.
\end{equation}
and refers to as \emph{Bernard-Duncan model}, with $r$ measuring the expansion rate. The great advantage of such an approach is that $a(\eta)$ becomes constant for $\eta\to\pm\infty$, fulfilling the requirements needed to use the prescription reported in Sec.~\ref{sec3}.

The study of the particle production is performed by considering $3$ different couplings $\zeta$ between the scalar field and the non-metricity scalar $Q$ (see the Lagrangian density in Eq.~\eqref{ansatz lagrangian STEGR}), i.e. $\zeta=0$ (minimal coupling), $\zeta=1/3$ (nullifying the term $\propto H^2$ in the right-hand side of Eq.~\eqref{potential always valid}) and $\zeta=2/3$. In this way, considering Eq.~\eqref{BD model}, the condition $|U(\eta)|\ll\omega_{in}^2$ validating the perturbative approach in Sec.~\ref{ssec4.2}, is satisfied if $1-a_{in}^2\ll\omega_{in}^2/m^2$ and $r^2\ll\omega_{in}^2$.

To satisfy the first condition, we may consider the production of massless particles. Bearing this request in mind, perturbative condition turns out to be immediately fulfilled. So, having $m=0$, since $k=\omega_{in}=\omega$, the particle production, in Eq.~\eqref{particle production perturbative}, becomes $N_{\omega}$, conventionally fixing $\omega=0.1$ and $r=0.01$.

Accordingly, we can find $p(\eta)$ and $\rho(\eta)$ by putting Eq.~\eqref{BD model} into Eq.~\eqref{friedmann equations modified gravity} when $f(Q)=Q$ and, then, from Eq.~\eqref{potential modified gravity}, we get $\max_{\eta}|U(\eta)|<5\cdot10^{-5}$, so that $|U(\eta)|\ll\omega^2$ is satisfied.
\begin{table}[h!]
\centering
\begin{tabular}{ |c|c|c|c| }
\hline
\hline
 $\zeta$ & $N_{\omega=0.1}$ & $C(E=10)$ & $\mathcal{Q}$ \\
 \hline
 \hline
 $0$ & $1.5831\cdot10^{-25}$ & $4.8345$ & $82.3854$\\
 \hline
 $1/3$ & $1.5833\cdot10^{-25}$ & $4.8345$ & $82.3853$\\
 \hline
 $2/3$ & $4.7439\cdot10^{-25}$ & $4.8345$ & $80.8021$\\
 \hline
 \hline
\end{tabular}\caption{In this table, the production of massless particles $N_{\omega=0.1}$ (Eq.~\eqref{particle production perturbative}), the constrained classical capacity $C(E=10)$ (Eq.~\eqref{constrained classical capacity ce}) and the quantum capacity $\mathcal{Q}$ (Eq.~\eqref{quantum capacity ce}) are reported for a Bernand-Duncan cosmological expansion \eqref{BD model}, with $r=0.01$ and $a_{in}=0.5$, for different values of the coupling $\zeta$ between the scalar field and the non-metricity scalar $Q$. The small cosmological particle production $N_{\omega=0.1}$ has practically no influence on the amount of classical information $C(E=10)$ a one-mode Gaussian state could preserve. On the contrary, the maximum amount of quantum information reliably stored by a Gaussian state $\mathcal{Q}$ is affected by the cosmological particle production $N_{\omega=0.1}$ even if very small.}
\label{table: unmodified case}
\end{table}
In so doing, we can calculate the particle production numerically from Eq.~\eqref{particle production perturbative} and the classical and quantum capacity by putting $N_k=N_{\omega}$ in Eqs.~\eqref{constrained classical capacity ce} and \eqref{quantum capacity ce}, respectively.

The corresponding numerical outcomes have been reported in table \ref{table: unmodified case}.

Switching to the classical capacity, if $N_\omega\ll 1$ and $E\gg N_\omega$, from Eq.~\eqref{constrained classical capacity ce}, we immediately have
\begin{equation}
    C(E)\sim h\left(\frac{1}{2}+E\right)+N_\omega\log N_\omega\,.
\end{equation}
As a consequence, as long as the particle production is very low, the classical capacity is practically unaffected by it\footnote{This is no more true if the number of particles used to encode the classical message $E$ is comparable to $N_\omega$. However, an extremely low $E$ leads always to a negligible classical capacity regardless $N_\omega$. For this reason, this case is not taken into account.}, as shown in the table \ref{table: unmodified case}.

The situation appears quite different for the quantum capacity. Indeed, from Tab. \ref{table: unmodified case}, we see that $\mathcal{Q}$ slightly  changes, for different $N_\omega$, even if $N_\omega\ll1$.

\subsection{Power-law modification of non-metric gravity}

We now see how the particle production and the capacities reported in Tab. \ref{table: unmodified case} change by considering a modified gravity theory $f(Q)$. Namely, we consider the power-law gravity modification in Eq.~\eqref{power-law modification}, where we conventionally fix the parameter $A$ as $A\sim Q$, and, in particular, we select $A=\max_\eta\{2\rho(\eta)\}\sim 1.418\cdot 10^{-4}$.

\begin{figure}[!htbp]
    \centering
    \includegraphics[scale=0.35]{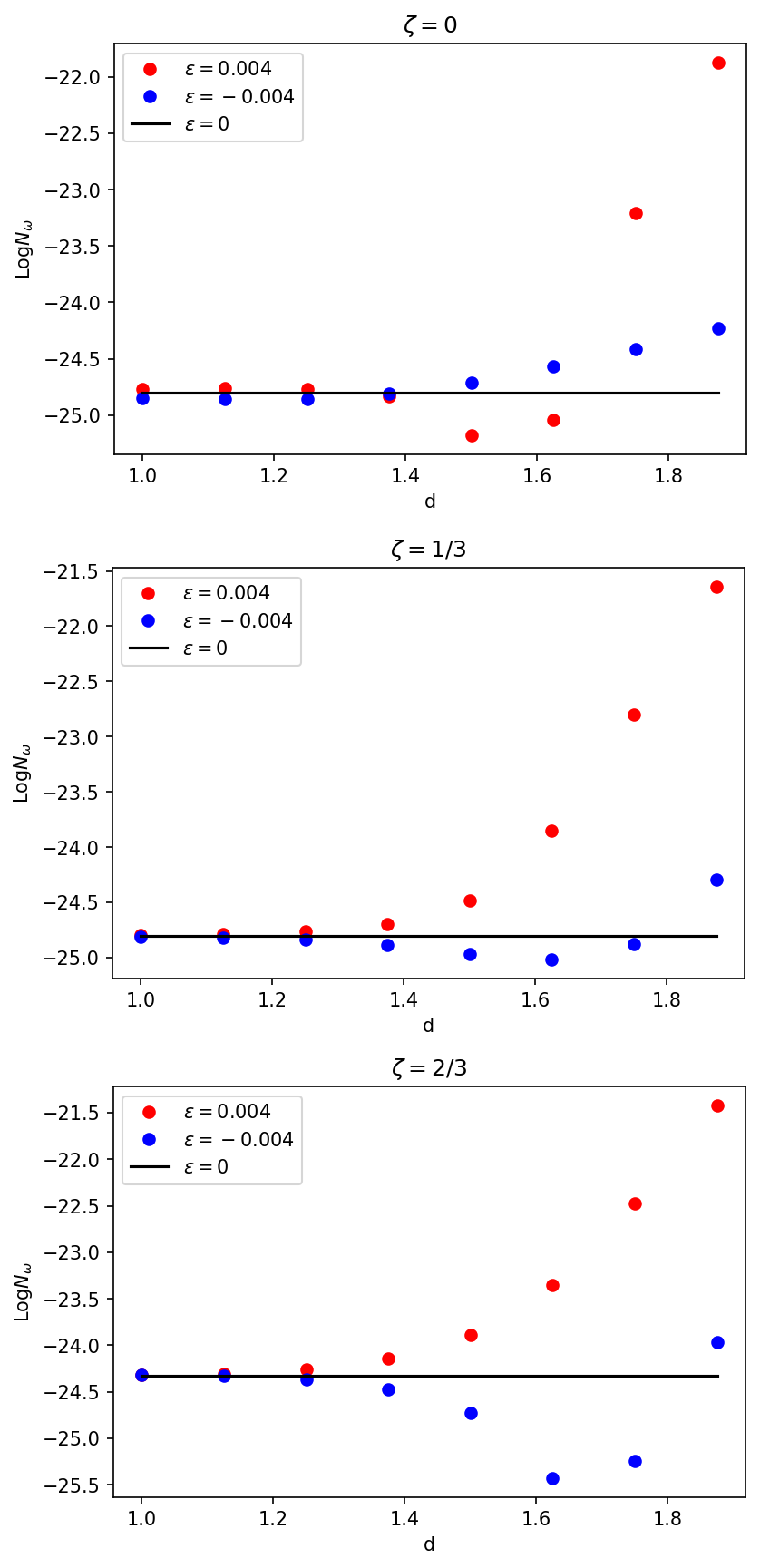}
    \caption{Number of massless scalar particles $N_\omega$, with frequency $\omega=0.1$, produced under a cosmological expansion given by the scale factor \eqref{BD model} (with $a_{in}=0.5$ and $r=0.01$) vs the parameter $d$ of the modified gravity theory $f(Q)$ from Eq.~\eqref{power-law modification}, where $A\simeq1.418\cdot10^{-4}$. The particle production $N_\omega$ is shown for different values of the coupling $\zeta$ between the particles' field and the non-metricity scalar $Q=6H^2$, and for different values of $\epsilon$, another parameter of the modification $f(Q)$ from Eq.~\eqref{power-law modification}.}
    \label{fig: particle production}
\end{figure}

With these parameters, the particle production was computed numerically inserting Eq.~\eqref{potential modified gravity} into Eq.~\eqref{particle production perturbative}. The results are shown in Fig.~\ref{fig: particle production} for different values of $\zeta$, $d$ and $\epsilon$ (the unmodified case, from Eq.~\eqref{power-law modification} corresponds to $\epsilon=0$).

As expected, the behaviour of the particle production is different depending on the sign of $\epsilon$. In case $\zeta=1/3$ or $\zeta=2/3$ the particle production $N_\omega$ is lower than the one obtained in general relativity only if $\epsilon<0$ and when $d$ assumes values in a bounded interval. Otherwise, if $\epsilon>0$, the particle production $N_\omega$ always increases with respect to the unmodified case $\epsilon=0$.

An opposite tendency can be observed from Fig.~\ref{fig: particle production} in the minimal coupling case $\zeta=0$, where the particle production $N_\omega$ is minimized when $\epsilon>0$. If $\epsilon<0$, we still have a decreasing of the particle production w.r.t. the unmodified case, but this is much smaller than the one get when $\epsilon>0$.

Since we showed in table \ref{table: unmodified case} that classical capacity is unaffected by a very low production of particles (see Table \ref{table: unmodified case}), we now focus on the quantum capacity only. Then, we analyze how much the quantum capacities, reported in Tab. \ref{table: unmodified case}, depart from GR, through extending it by virtue of $f(Q)$, see Fig.~\ref{fig: particle production}. For each coupling $\zeta$ the highest quantum capacity occurs when $d=d_{min}$, minimizing the particle production to $\underline{N}_\omega$. Then, we can quantify the maximum gain of quantum capacity through
\begin{equation}\label{gain of quantum capacity}
    \Delta\mathcal{Q}=\frac{\mathcal{Q}(N_\omega=\underline{N}_\omega)-\mathcal{Q}(\epsilon=0)}{\mathcal{Q}(\epsilon=0)}\,.
\end{equation}
The values of $\Delta\mathcal{Q}$ are reported in Tab. \ref{table: modified case}, as well as the numerically computed values of $d_{min}$.
\begin{table}[h!]
\centering
\begin{tabular}{ |c|c|c|c| }
\hline
\hline
 $\zeta$ & $\epsilon$ & $d_{min}$ & $\Delta \mathcal{Q}$ \\
 \hline
 \hline
 $0$ & $0.004$ & $1.59$ & $4.23\%$\\
 \hline
 $1/3$ & $-0.004$ & $1.61$ & $0.89\%$\\
 \hline
 $2/3$ & $-0.004$ & $1.69$ & $7.76\%$\\
 \hline
 \hline
\end{tabular}\caption{In this table we report, for each value of the coupling $\zeta$, the maximum gain of quantum capacity $\Delta\mathcal{Q}$ (from Eq.~\eqref{gain of quantum capacity}) achievable with the modified gravity theory \eqref{power-law modification} by bounding $|\epsilon|=0.004$. In the second and third columns we report respectively the values of the parameters $\epsilon$ and $d$ of the power law modification needed to reach the gain $\Delta\mathcal{Q}$.}\label{table: modified case}
\end{table}
In general, we proved that, for each coupling $\zeta$, there is a polynomial $f(Q)$ model, with parameters established in table \ref{table: modified case}, preserving quantum information from remote past better than general relativity. With these theories, we are able to reach a modest increasing on the conserved quantum information (up to $4.23\%$ in case of minimal coupling $\zeta=0$). Moreover, these modified theories also preserve the GR gravitational coupling in the far future, so that local observations would not be disrupted by them.

\section{Discussion and Conclusions}\label{conclusioni}

In this work, we investigated the maximum amount of classical and quantum information that a one-mode Gaussian state can reliably preserve from the remote past through cosmological expansion. To achieve this, we developed a general method based on quantum communication theory, recognizing the Bogoliubov transformations of one-mode bosonic states as quantum channels. We analyzed their properties and capacities in terms of the Bogoliubov coefficients, i.e., adopting a method also useful for studying how entanglement entropy evolves from the remote past to the distant future, as reported in Appendices \ref{appendixA} and \ref{appendixB}.

Particularly, for a generic cosmological expansion in a homogeneous and isotropic universe, the resulting particle production is shown to be responsible for amplifying bosonic signals that originate from the remote past. Consequently, the classical and quantum information, encoded in these signals, degrades due to the cosmological particle production. We focused on the cosmological scenario that best preserves information from the remote past, finding that it represents the one with the least amount of produced particles.

As reported in detail in Appendix \ref{appendixB}, we demonstrated that entanglement, originating from the remote past, is most significantly enhanced by increasing cosmological particle production. We showed that this result extends the well-known phenomenon of entanglement creation from vacuum by a cosmological expansion for each initial maximally entangled state. In this context, it is shown that entanglement from the remote past always increases when the initial number of entangled particles is small. However, if the initial entanglement entropy is sufficiently high, this enhancement is not always guaranteed. Indeed, for certain values of the phase of the Bogoliubov coefficients, cosmological particle production may even lead to entanglement degradation.

In this respect, we also investigated how modified theories of gravity that lead to different results. We thus focused on $f(Q)$ theories of gravity, in which analytical functions of non-metricity are involved. Precisely, we circumscribed our analysis to those theories that can decrease particle production. To this end, we showed that very simple power-law modifications, with appropriate set of free parameters matching local observation, can optimize the preservation of classical and quantum information stored in bosons from remote past. Notably, while classical information preserved in a system remains unaffected by small particle production, the quantum information in the same system can be significantly compromised even with minimal cosmological particle production.

By adjusting the parameters of the power-law gravity modification, the amount of preserved quantum information can increase up to $4.23\%$ compared to the prediction of GR, in the case of minimal coupling ($\zeta=0$). By increasing the coupling to $\zeta=2/3$, $f(Q)$ theories  can increase the maximum amount of conserved quantum information, up to $\sim8\%$.

Summing up, the results here obtained may therefore open the possibility of detecting cosmological particle production through the study of the degradation of qubit information content. Consequently, the ability to achieve cosmological particle production, including through analog gravity systems \cite{Jain_2007}, offers a pathway to distinguish between modified theories of gravity and the coupling between fields and curvature (or non-metricity). In conclusion, cosmological particle production serves as a crucial tool for understanding the evolutionary history of our universe, particularly during the inflationary epoch, which is believed to have generated a substantial quantity of particles \cite{Barnaby_2009, Kolb_2023, Cembranos2023}. Recent studies have explored models linking dark matter to this particle production process \cite{Li_2019, Li_2020, Safari_2022}. Within this framework, the reduction in communication capacities due to particle production holds profound implications. It is plausible that the dark particles produced during inflation could obscure information regarding particles from the remote past. As a result, this research, along with related studies focusing on fermions \cite{Mancini_2014}, may establish fundamental limits on our ability to recover information about primordial particles.

\section*{Acknowledgements}

This paper is based upon work from COST Action CA21136
Addressing observational tensions in cosmology with systematics and fundamental physics (CosmoVerse) supported by COST (European Cooperation in Science and Technology). SC acknowledges the Istituto Italiano di Fisica Nucleare (INFN) iniziative specifiche QGSKY and MOONLIGHT2. OL acknowledges hospitality to the Al-Farabi Kazakh National University during the time in which this paper has been thought. He is also grateful to INAF, National Institute of Astrophysics, for the support and, particularly, to Roberto della Ceca and Luigi Guzzo. SM acknowledges financial support from “PNRR MUR project PE0000023-NQSTI”.

\appendix

\section{Entropy and entanglement of bosonic Gaussian states}\label{appendixA}
Here we provide further mathematical information on the BGSs studied in Sec.~\ref{sec2}, focusing on their Von Neumann entropy and on their entanglement entropy. A two-modes Gaussian state (modes $i,j$) can be written, from Eqs.~\eqref{completeCovMatrix}, \eqref{one-mode cov matrix} and \eqref{correlation matrix}, as
\begin{widetext}
\begin{equation}\label{two mode covariance matrix}
    \sigma=\left(\begin{matrix}
        \frac{1}{2}+n_i+\Re m_i&\Im m_i&\Re\gamma_{ij}+\Re\chi_{ij}&\Im\gamma_{ij}+\Im\chi_{ij}\\\Im m_i&\frac{1}{2}+n_i-\Re m_i&-\Im\gamma_{ij}+\Im\chi_{ij}&\Re\gamma_{ij}-\Re\chi_{ij}\\
        \Re\gamma_{ij}+\Re\chi_{ij}&-\Im\gamma_{ij}+\Im\chi_{ij}&\frac{1}{2}+n_j+\Re m_j&\Im m_j\\\Im\gamma_{ij}+\Im\chi_{ij}&\Re\gamma_{ij}-\Re\chi_{ij}&\Im m_j&\frac{1}{2}+n_j-\Re m_j
    \end{matrix}\right)\,,
\end{equation}\normalsize
\end{widetext}
where $\gamma_{ij}=\langle a_i^\dagger a_j\rangle$, $\chi_{ij}=\langle a_ia_j\rangle$. The calculation of the symplectic eigenvalues of $4\times4$ definite positive matrices is provided in literature (see e.g. Ref.~\cite{Laurat_2005}) and they are
\begin{equation}\label{two mode symplectic eigenvalues}
    \nu_{\pm}=\frac{1}{\sqrt{2}}\sqrt{\Gamma\pm\sqrt{\Gamma^2-4\det\sigma}}\,,
\end{equation}
where we defined the parameter $\Gamma\coloneqq\nu_i^2+\nu_j^2+2\left(|\gamma_{ij}|^2-|\chi_{ij}|^2\right)$ with $\nu_i$ ($\nu_j$) the symplectic eigenvalue of the covariance matrix of the subsystem state $\sigma_{ii}$ ($\sigma_{jj}$), given by
\begin{equation}
    \nu_i\coloneqq\sqrt{\det\sigma_{ii}}
\end{equation}
The Von Neumann entropy of the two mode Gaussian state is then
\begin{equation}\label{two modes entropy}
    \mathcal{S}(\sigma)=h(\nu_-)+h(\nu_+)\,.
\end{equation}
A maximally entangled state is a composite pure state with zero entropy and where the entropy of one of its subsystem states - also called \textit{entanglement entropy} $\mathcal{S}_E$ - is maximized. Fixing the average number of particles in each subsystem state $n_i=n_j$, a maximally entangled two-modes Gaussian state is provided when $m_i=0$, $\gamma_{ij}=0$ and $|\chi_{ij}|=\sqrt{n_i(n_i+1)}$. The covariance matrix of this state is\begin{widetext}
\begin{equation}\label{input cov matrix}
    \sigma=\left(
        \begin{matrix}
            \frac{1}{2}+n_i&0&\sqrt{n_i(n_i+1)}\cos\theta&\sqrt{n_i(n_i+1)}\sin\theta\\
            0&\frac{1}{2}+n_i&\sqrt{n_i(n_i+1)}\sin\theta&-\sqrt{n_i(n_i+1)}\cos\theta\\
            \sqrt{n_i(n_i+1)}\cos\theta&\sqrt{n_i(n_i+1)}\sin\theta&\frac{1}{2}+n_i&0\\
            \sqrt{n_i(n_i+1)}\sin\theta&-\sqrt{n_i(n_i+1)}\cos\theta&0&\frac{1}{2}+n_i
        \end{matrix}
    \right)\,,
\end{equation}\end{widetext}
where $\theta$ is the phase of $\chi_{ij}$. Its entanglement entropy is
\begin{equation}
    \mathcal{S}_E(\sigma)=h\left(\frac{1}{2}+n_i\right)\,,
\end{equation}
and it is proportional on the average number of entangled particles $n_i$.

\section{Entanglement preservation after a cosmological expansion}\label{appendixB}
From the literature (see e.g. \cite{Ball_2006,Fuentes_2010,Mart_n_Mart_nez_2012}), it is well known that the particles created from vacuum, with momentum $\mathbf{k}$ by a cosmological expansion are maximally entangled with the particle created from vacuum with momentum $-\mathbf{k}$. By virtue of this, one may think that particle production reinforces any pre-existing entanglement. We show in this appendix that this is not the case. Namely, there can be states whose entanglement is degraded by particle production.

From now on, calling $i$ the mode relative to the momentum $\mathbf{k}$, we refer with $-i$ to the mode relative to its opposite momentum $-\mathbf{k}$. We then simplify Eqs.~\eqref{n tilde}, \eqref{m tilde}, \eqref{gamma tilde} and \eqref{chi tilde} using the Bogoliubov coefficients \eqref{alphasimp} and \eqref{betasimp}, getting
\begin{equation}\label{n tilde cosmological exp}
    \tilde{n}_i=|\alpha_i|^2n_i+|\beta_i|^2(1+n_{-i})-2\Re\left(\alpha_i^\ast\beta_i\chi_{i,-i}\right)\,;
\end{equation}
\begin{equation}\label{m tilde cosmological exp}
    \tilde{m}_i=\alpha_i^{\ast2}m_i+\beta_i^{\ast2}m_{-i}-2\alpha_i\beta_i^\ast\gamma_{i,-i}^\ast\,.
\end{equation}
\begin{equation}\label{gamma tilde cosmological exp}
    \Tilde{\gamma}_{i,-i}=-\alpha_i\beta_i^\ast m_i^\ast-\alpha_i^\ast\beta_i m_{-i}+(|\alpha_i|^2+|\beta_i|^2)\gamma_{i,-i}\,;
\end{equation}
\begin{equation}\label{chi tilde cosmological expansion}
    \tilde{\chi}_{i,-i}=-\alpha_i^\ast\beta_i^\ast(1+n_i)-\beta_i^\ast\alpha_i^\ast n_{-i}+(\alpha_i^{\ast 2}+\beta_i^{\ast 2})\chi_{i,-i}\,.
\end{equation}
As specified in Eq.~\eqref{input cov matrix}, a maximally entangled two-modes Gaussian state with number of particles in the subsystem states $n_i=n_{-i}$ is characterized by the covariance matrix \eqref{two mode covariance matrix} with $m_{\pm i}=0$, $\gamma_{i,-i}=0$ and $\chi_{i,-i}=\sqrt{n_i(n_i+1)}e^{i\theta}$ for an arbitrary $\theta$. The state after the cosmological expansion, called $\tilde{\sigma}$, is again characterized by the two-modes state \eqref{two mode covariance matrix} with $\tilde{n}_{\pm i}$, $\Tilde{m}_{\pm i}$, $\Tilde{\gamma}_{i,-i}$ and $\tilde{\chi}_{i,-i}$ replacing $n_{\pm i}$, $m_{\pm i}$, $\gamma_{i,-i}$ and $\chi_{i,-i}$, respectively. Using Eqs.~\eqref{n tilde cosmological exp}, \eqref{m tilde cosmological exp}, \eqref{gamma tilde cosmological exp} and \eqref{chi tilde cosmological expansion}, we get
\begin{equation}\label{output n}
    \tilde{n}_{\pm i}=(1+2N_i)n_i+N_i-2\Re(\alpha^\ast_i\beta_i\chi_{i,-i})\,;
\end{equation}
\begin{equation}\label{output m}
    \Tilde{m}_i=0\,;
\end{equation}
\begin{equation}\label{output gamma}
    \Tilde{\gamma}_{\pm i,\mp i}=0\,;
\end{equation}
\begin{equation}\label{output chi}
    \tilde{\chi}_{\pm i,\mp i}=-\alpha_i^\ast\beta_i^\ast(1+2n_i)+\alpha_i^{\ast 2}\chi_{i,-i}+\beta_i^{\ast 2}\chi_{i,-i}^\ast\,.
\end{equation}
If we study the symplectic eigenvalues of the state \eqref{two mode symplectic eigenvalues} using the parameters \eqref{output n}, \eqref{output m}, \eqref{output gamma} and \eqref{output chi} we get that they are both $1/2$. As a consequence, from Eq.~\eqref{two modes entropy}, the entropy of the cosmological expansion output $\Tilde{\sigma}$ is zero. This means that $\tilde{\sigma}$ is still a maximally entangled state and its entanglement entropy is
\begin{align}\label{entanglement entropy final}
    \mathcal{S}_E(\tilde{\sigma})&=h\left(\frac{1}{2}+\tilde{n}_i\right)
    =h\left(\frac{1}{2}+n_i+N_i(1+2n_i)\right.\nonumber\\&\left.-2\sqrt{n_i(n_i+1)}\sqrt{N_i(N_i+1)}\cos(\varphi+\theta)\right)\,,
\end{align}
where we defined $\varphi$ from $\alpha_i^\ast\beta_i=\sqrt{N_i(N_i+1)}e^{i\varphi}$. We can immediately verify that the result is consistent with the literature \cite{Ball_2006,Fuentes_2010,Mart_n_Mart_nez_2012} in the vacuum case, i.e. for $n_i=0$.
To have the enhancement of entanglement, we need $\mathcal{S}_E(\sigma)<\mathcal{S}_E(\Tilde{\sigma})$, occurring when $n_i<\Tilde{n}_i$, i.e. when
\begin{equation}
    \label{condition enhancement general}
    \cos(\theta+\varphi)<\frac{1+2n_i}{2\sqrt{n_i(n_i+1)}}\sqrt{\frac{N_i}{N_i+1}}\,.
\end{equation}
The condition \eqref{condition enhancement general} is strongly dependent on the parameter $\theta$ of the initially entangled state \eqref{input cov matrix} (assuming $\varphi$ fixed by the cosmic expansion). Namely, in terms of $\theta$, the condition for the entanglement enhancement becomes
\begin{align}
    \theta\notin &\left(\arccos\left(\frac{1+2n_i}{2\sqrt{n_i(n_i+1)}}\sqrt{\frac{N_i}{N_i+1}}\right)-\varphi,\right.\nonumber\\&\left.-\arccos\left(\frac{1+2n_i}{2\sqrt{n_i(n_i+1)}}\sqrt{\frac{N_i}{N_i+1}}\right)-\varphi\right)\,.\label{condition enhancement 2}
\end{align}
Instead, if $\theta$ belongs to the interval specified in Eq.~\eqref{condition enhancement 2}, the entanglement in the far future is less than the one at remote past. Namely, the cosmological expansion causes \textit{entanglement degradation}. In this case, comparing the initial entanglement entropy with the final one, in Eq.~\eqref{entanglement entropy final}, we can easily seen that the amount of entanglement degraded vanishes for $N_i\to0$. Hence, the scenario with less particle production, researched throughout the paper, is the one reducing the degradation of entanglement if the condition \eqref{condition enhancement 2} is not satisfied.

For the sake of completeness, it is worth noticing that, if the particle production is enough high, the entanglement of a state from the remote past is always enhanced by the cosmological expansion, regardless the value of $\theta$. This happens when the right hand side of Eq.~\eqref{condition enhancement general} is greater than $1$, namely when
\begin{equation}\label{enhancement guaranteed}
    N_i>4n_i+4n_i^2.
\end{equation}
This is consistent with the fact that, starting from the vacuum $n_i=0$, the creation of entanglement is guaranteed by the particle production. To conclude, if the aim is maximizing the entanglement at the far future, a high particle production scenario is preferable, so that Eq.~\eqref{enhancement guaranteed} is satisfied. However, if the aim is to better preserve entanglement from the remote past, for each number of initially entangled particles $n_i$, the preferable scenario is the one minimizing the particle production.


\begin{thebibliography}{10}

\bibitem{shannon}
C.~E. Shannon.
\newblock A mathematical theory of communication.
\newblock {\em The Bell System Technical Journal}, 27(3):379--423, July 1948.

\bibitem{Gatenby2007}
R.~Gatenby and R.~Frieden.
\newblock Information theory in living systems, methods, applications, and challenges.
\newblock {\em Bulletin of mathematical biology}, 69(2):635--57, March 2007.

\bibitem{mancini2019quantum}
S.~Mancini and A.~Winter.
\newblock {\em A Quantum Leap in Information Theory}.
\newblock World Scientific, 2019.

\bibitem{BEKENSTEIN_1990}
J.~D. Bekenstein and M.~Schiffer.
\newblock Quantum limitations on the storage and transmission of information.
\newblock {\em Int.~J.~Mod.~Phys.~C}, 01(04):355--422, December 1990.

\bibitem{bekenstein2003information}
J.~D. Bekenstein.
\newblock Information in the holographic universe.
\newblock {\em Scientific American}, 289(2):58--65, August 2003.

\bibitem{Sidhu_2021}
J.~S.~Sidhu et~Al.
\newblock Advances in space quantum communications.
\newblock {\em {IET} Quantum Communication}, 2(4):182--217, July 2021.

\bibitem{DOLGOV20031}
A.~D. Dolgov and M.~Kawasaki.
\newblock Can modified gravity explain accelerated cosmic expansion?
\newblock {\em Phys.~Lett.~B}, 573:1--4, October 2003.

\bibitem{Faraoni2007}
V.~Faraoni and A.~Jacques.
\newblock Cosmological expansion and local physics.
\newblock {\em Phys.~Rev.~D}, 76(6):063510, September 2007.

\bibitem{Mann_2012}
R.~B. Mann and T.~C. Ralph.
\newblock Relativistic quantum information.
\newblock {\em Class.~Quantum Grav.}, 29(22):220301, November 2012.

\bibitem{Bruckner2014}
{\v{C}}.~Brukner.
\newblock {Quantum causality}.
\newblock {\em Nature Physics}, 10(4):259--263, April 2014.

\bibitem{DeRamon2023}
J.~de~Ram\'on, M.~Papageorgiou, and E.~Mart\'{\i}n-Mart\'{\i}nez.
\newblock Causality and signalling in noncompact detector-field interactions.
\newblock {\em Phys.~Rev.~D}, 108(4):045015, August 2023.

\bibitem{braunstein2007quantum}
S.~L. Braunstein and A.~K. Pati.
\newblock Quantum information cannot be completely hidden in correlations: implications for the black-hole information paradox.
\newblock {\em Phys.~Rev.~Lett.}, 98(8):080502, February 2007.

\bibitem{chen2022quantum}
B.~Chen, B.~Czech, and Z.~Wang.
\newblock Quantum information in holographic duality.
\newblock {\em Reports on Progress in Physics}, 85(4):046001, March 2022.

\bibitem{Capozziello:2024ucm}
Salvatore Capozziello, Silvia De~Bianchi, and Emmanuele Battista.
\newblock {Avoiding singularities in Lorentzian-Euclidean black holes: The role of~atemporality}.
\newblock {\em Phys. Rev. D}, 109(10):104060, May 2024.

\bibitem{van2020quantum}
T.~W. Van De~Kamp, R.~J. Marshman, S.~Bose, and A.~Mazumdar.
\newblock Quantum gravity witness via entanglement of masses: Casimir screening.
\newblock {\em Phys.~Rev.~A}, 102(6):062807, December 2020.

\bibitem{Christodoulou2023}
M.~Christodoulou et~Al.
\newblock Locally mediated entanglement in linearized quantum gravity.
\newblock {\em Phys.~Rev.~Lett.}, 130(10):100202, March 2023.

\bibitem{bruschi2013robustness}
D.~E. Bruschi, N.~Friis, I.~Fuentes, and S.~Weinfurtner.
\newblock On the robustness of entanglement in analogue gravity systems.
\newblock {\em New Journal of Physics}, 15(11):113016, November 2013.

\bibitem{jacquet2020next}
M.~Jacquet, S.~Weinfurtner, and F.~Koenig.
\newblock The next generation of analogue gravity experiments.
\newblock {\em Philosophical Transactions A}, 378(2177):20190239, July 2020.

\bibitem{Birrell:1982ix}
N.~D. Birrell and P.~C.~W. Davies.
\newblock {\em {Quantum Fields in Curved Space}}.
\newblock Cambridge Monographs on Mathematical Physics. Cambridge Univ. Press, Cambridge, UK, 1984.

\bibitem{parker_toms_2009}
L.~Parker and D.~Toms.
\newblock {\em Quantum Field Theory in Curved Spacetime: Quantized Fields and Gravity}.
\newblock Cambridge Monographs on Mathematical Physics. Cambridge University Press, 2009.

\bibitem{anastopoulos2023quantum}
C.~Anastopoulos, B.~Hu, and K.~Savvidou.
\newblock Quantum field theory based quantum information: Measurements and correlations.
\newblock {\em Ann.~Phys.}, 450:169239, March 2023.

\bibitem{perche2023role}
T.~R. Perche and E.~Mart{\'\i}n-Mart{\'\i}nez.
\newblock Role of quantum degrees of freedom of relativistic fields in quantum information protocols.
\newblock {\em Phys.~Rev.~A}, 107(4):042612, April 2023.

\bibitem{Unruh1976}
W.~G. Unruh.
\newblock Notes on black-hole evaporation.
\newblock {\em Phys.~Rev.~D}, 14(4):870--892, August 1976.

\bibitem{Unruh1984}
W.~G. Unruh and R.~M. Wald.
\newblock What happens when an accelerating observer detects a rindler particle.
\newblock {\em Phys.~Rev.~D}, 29(6):1047--1056, March 1984.

\bibitem{Hu_2012}
B.~L. Hu, S.~Lin, and J.~Louko.
\newblock Relativistic quantum information in detectors{\textendash}field interactions.
\newblock {\em Class.~Quantum Grav.}, 29(22):224005, October 2012.

\bibitem{Brown_2013}
E.~G. Brown, E.~Mart{\'{\i}}n-Mart{\'{\i}}nez, N.~C. Menicucci, and R.~B. Mann.
\newblock Detectors for probing relativistic quantum physics beyond perturbation theory.
\newblock {\em Phys.~Rev.~D}, 87(8):084062, April 2013.

\bibitem{Tjoa_2022}
E.~Tjoa and K.~Gallock-Yoshimura.
\newblock Channel capacity of relativistic quantum communication with rapid interaction.
\newblock {\em Phys.~Rev.~D}, 105(8):085011, April 2022.

\bibitem{Lapponi_2023}
A.~Lapponi, D.~Moustos, D.~E. Bruschi, and S.~Mancini.
\newblock Relativistic quantum communication between harmonic oscillator detectors.
\newblock {\em Phys.~Rev.~D}, 107(12):125010, June 2023.

\bibitem{lapponi2024making}
A.~Lapponi, J.~Louko, and S.~Mancini.
\newblock Making two particle detectors in flat spacetime communicate quantumly.
\newblock {\em arXiv:2404.01880}, April 2024.
\newblock Accepted in Phys.~Rev.~D in June 2024.

\bibitem{Br_dler_2014}
K.~Br{\'{a}}dler and C.~Adami.
\newblock The capacity of black holes to transmit quantum information.
\newblock {\em J.~High Energy Phys.}, 2014(095), May 2014.

\bibitem{Mancini_2014}
S.~Mancini, R.~Pierini, and M.~M. Wilde.
\newblock Preserving information from the beginning to the end of time in a robertson{\textendash}walker spacetime.
\newblock {\em New Journal of Physics}, 16(12):123049, December 2014.

\bibitem{Gianfelici_2017}
G.~Gianfelici and S.~Mancini.
\newblock Quantum channels from reflections on moving mirrors.
\newblock {\em Scientific Reports}, 7(1):15747, November 2017.

\bibitem{Good_2021}
M.~R.{\hspace{0.167em} }R. Good, A.~Lapponi, O.~Luongo, and S.~Mancini.
\newblock Quantum communication through a partially reflecting accelerating mirror.
\newblock {\em Phys.~Rev.~D}, 104(10):105020, November 2021.

\bibitem{Jaynes1963}
E.~T. Jaynes and F.~W. Cummings.
\newblock Comparison of quantum and semiclassical radiation theories with application to the beam maser.
\newblock {\em Proceedings of the IEEE}, 51(1):89--109, February 1963.

\bibitem{Bruschi_2013}
D.~E. Bruschi, A.~R. Lee, and I.~Fuentes.
\newblock Time evolution techniques for detectors in relativistic quantum information.
\newblock {\em J.~Phys.~A: Mathematical and Theoretical}, 46(16):165303, April 2013.

\bibitem{Ford_2021}
L.~H. Ford.
\newblock Cosmological particle production: a review.
\newblock {\em Reports on Progress in Physics}, 84(11):116901, October 2021.

\bibitem{Ball_2006}
J.~L. Ball, I.~Fuentes-Schuller, and F.~P. Schuller.
\newblock Entanglement in an expanding spacetime.
\newblock {\em Phys.~Lett.~A}, 359(6):550--554, December 2006.

\bibitem{Fuentes_2010}
I.~Fuentes, R.~B. Mann, E.~Mart{\'{\i}}n-Mart{\'{\i}}nez, and S.~Moradi.
\newblock Entanglement of {Dirac} fields in an expanding spacetime.
\newblock {\em Phys.~Rev.~D}, 82(4):045030, August 2010.

\bibitem{Mart_n_Mart_nez_2012}
E.~Mart{\'{\i}}n-Mart{\'{\i}}nez and N.~C. Menicucci.
\newblock Cosmological quantum entanglement.
\newblock {\em Class.~Quantum Grav.}, 29(22):224003, October 2012.

\bibitem{serafini2017quantum}
A.~Serafini.
\newblock {\em Quantum Continuous Variables: A Primer of Theoretical Methods}.
\newblock CRC Press, 2017.

\bibitem{2019ACO}
A century of correct predictions.
\newblock {\em Nature Physics}, 15:415, May 2019.

\bibitem{nojiri2007introduction}
S.~Nojiri and S.~D. Odintsov.
\newblock Introduction to modified gravity and gravitational alternative for dark energy.
\newblock {\em Int.~J.~Geom.~Methods Mod.~Phys.}, 4(01):115--145, February 2007.

\bibitem{Capozziello_2007}
S.~Capozziello and M.~Francaviglia.
\newblock Extended theories of gravity and their cosmological and astrophysical applications.
\newblock {\em General Relativity and Gravitation}, 40(2-3):357--420, December 2007.

\bibitem{De_Felice_2010}
A.~De Felice and S.~Tsujikawa.
\newblock $f({R})$ theories.
\newblock {\em Living Rev.~Relativ.}, 13(1), June 2010.

\bibitem{capozziello2011extended}
S.~Capozziello and M.~De~Laurentis.
\newblock Extended theories of gravity.
\newblock {\em Phys.~Rep.~}, 509(4-5):167--321, December 2011.

\bibitem{lombriser2012cluster}
L.~Lombriser et~Al.
\newblock Cluster density profiles as a test of modified gravity.
\newblock {\em Phys.~Rev.~D}, 85(10):102001, May 2012.

\bibitem{reina2023initial}
M.~Reina-Campos, A.~Sills, and G.~Bichon.
\newblock Initial sizes of star clusters: implications for cluster dissolution during galaxy evolution.
\newblock {\em Mon.~Not.~R.~Astron.~Soc.}, 524(1):968--980, September 2023.

\bibitem{nojiri2010future}
S.~Nojiri and S.~D. Odintsov.
\newblock Is the future universe singular: Dark matter versus modified gravity?
\newblock {\em Phys.~Lett.~B}, 686(1):44--48, March 2010.

\bibitem{davari2020testing}
Z.~Davari and S.~Rahvar.
\newblock Testing modified gravity ({MOG}) theory and dark matter model in {Milky} {Way} using the local observables.
\newblock {\em Mon.~Not.~R.~Astron.~Soc.}, 496(3):3502--3511, August 2020.

\bibitem{ADAK_2013}
M.~Adak, Ö. Sert, M.~Kalay, and M.~Sari.
\newblock Symmetric teleparallel gravity: some exact solutions and spinor couplings.
\newblock {\em Int.~J.~Mod.~Phys.~A}, 28(32):1350167, December 2013.

\bibitem{mandal2020cosmography}
S.~Mandal, D.~Wang, and PK~Sahoo.
\newblock Cosmography in $f({Q})$ gravity.
\newblock {\em Phys.~Rev.~D}, 102(12):124029, December 2020.

\bibitem{Jim_nez_2020}
J.~B. Jim{\'{e}}nez, L.~Heisenberg, T.~Koivisto, and S.~Pekar.
\newblock Cosmology in {$f({Q})$} geometry.
\newblock {\em Phys.~Rev.~D}, 101(10):103507, May 2020.

\bibitem{Heisenberg:2023lru}
Lavinia Heisenberg.
\newblock {Review on {$f(Q)$} gravity}.
\newblock {\em Phys. Rept.}, 1066:1--78, May 2024.

\bibitem{jimenez2019geometrical}
J.~Jimenez, L.~Heisenberg, and T.~Koivisto.
\newblock The geometrical trinity of gravity.
\newblock {\em Universe}, 5(7):173, July 2019.

\bibitem{Capozziello:2022zzh}
Salvatore Capozziello, Vittorio De~Falco, and Carmen Ferrara.
\newblock {Comparing equivalent gravities: common features and differences}.
\newblock {\em Eur. Phys. J. C}, 82(10):865, October 2022.

\bibitem{maurya2022anisotropic}
S.~K. Maurya, K.~N. Singh, S.~V. Lohakare, and B.~Mishra.
\newblock Anisotropic strange star model beyond standard maximum mass limit by gravitational decoupling in $f({Q})$ gravity.
\newblock {\em Fortschritte der Physik}, 70(11):2200061, September 2022.

\bibitem{bhar2023physical}
P.~Bhar, S.~Pradhan, A.~Malik, and P.~K. Sahoo.
\newblock Physical characteristics and maximum allowable mass of hybrid star in the context of $f({Q})$ gravity.
\newblock {\em Eur.~Phys.~J.~C}, 83(7):646, July 2023.

\bibitem{Capozziello:2024vix}
Salvatore Capozziello, Maurizio Capriolo, and Shin'ichi Nojiri.
\newblock {Gravitational waves in {$f(Q)$} non-metric gravity via geodesic deviation}.
\newblock {\em Phys. Lett. B}, 850:138510, March 2024.

\bibitem{Capozziello:2024jir}
Salvatore Capozziello and Maurizio Capriolo.
\newblock {Gravitational waves in {$f(Q)$} non-metric gravity without gauge fixing}.
\newblock {\em Phys. Dark Univ.}, 45:101548, June 2024.

\bibitem{ALBUQUERQUE2022100980}
I.~S. Albuquerque and N.~Frusciante.
\newblock A designer approach to $f({Q})$ gravity and cosmological implications.
\newblock {\em Phys.~Dark Universe}, 35:100980, March 2022.

\bibitem{Khyllep_2023}
W.~Khyllep, J.~Dutta, E.~N. Saridakis, and K.~Yesmakhanova.
\newblock Cosmology in $f({Q})$ gravity: A unified dynamical systems analysis of the background and perturbations.
\newblock {\em Phys.~Rev.~D}, 107(4):044022, February 2023.

\bibitem{capozziello2022model}
S.~Capozziello and R.~D'Agostino.
\newblock Model-independent reconstruction of $f({Q})$ non-metric gravity.
\newblock {\em Phys.~Lett.~B}, 832:137229, September 2022.

\bibitem{Nojiri:2024zab}
Shin'ichi Nojiri and S.~D. Odintsov.
\newblock {Well-defined {$f(Q)$} gravity, reconstruction of FLRW spacetime and unification of inflation with dark energy epoch}.
\newblock {\em Phys. Dark Univ.}, 45:101538, April 2024.

\bibitem{Ayuso2021}
I.~Ayuso, R.~Lazkoz, and V.~Salzano.
\newblock Observational constraints on cosmological solutions of $f({Q})$ theories.
\newblock {\em Phys.~Rev.~D}, 103(6):063505, March 2021.

\bibitem{Anagnostopoulos_2021}
F.~K. Anagnostopoulos, S.~Basilakos, and E.~N. Saridakis.
\newblock First evidence that non-metricity $f({Q})$ gravity could challenge {Lambda-CDM}.
\newblock {\em Phys.~Lett.~B}, 822(4):136634, November 2021.

\bibitem{Adesso_2014}
G.~Adesso, S.~Ragy, and A.~R. Lee.
\newblock Continuous variable quantum information: Gaussian states and beyond.
\newblock {\em Open Systems {\&} Information Dynamics}, 21(01n02):1440001, March 2014.

\bibitem{devetak2004capacity}
I.~Devetak and P.~W. Shor.
\newblock The capacity of a quantum channel for simultaneous transmission of classical and quantum information.
\newblock {\em Commun.~Math.~Phys.~}, 256:287--303, March 2003.

\bibitem{Caruso_2006}
F.~Caruso, V.~Giovannetti, and A.~S. Holevo.
\newblock One-mode bosonic gaussian channels: a full weak-degradability classification.
\newblock {\em New Journal of Physics}, 8(12):310--310, December 2006.

\bibitem{Pilyavets_2012}
O.~V. Pilyavets, C.~Lupo, and S.~Mancini.
\newblock Methods for estimating capacities and rates of {Gaussian} quantum channels.
\newblock {\em {IEEE} Trans.~Inf.~Theory}, 58(9):6126--6164, September 2012.

\bibitem{Br_dler_2015}
K.~Br{\'{a}}dler.
\newblock Coherent information of one-mode {Gaussian} channels{\textemdash}the general case of non-zero added classical noise.
\newblock {\em J.~Phys.~A: Mathematical and Theoretical}, 48(12):125301, March 2015.

\bibitem{BERNARD1977201}
C.~Bernard and A.~Duncan.
\newblock Regularization and renormalization of quantum field theory in curved space-time.
\newblock {\em Ann.~Phys.}, 107(1):201--221, September 1977.

\bibitem{Ford1987}
L.~H. Ford.
\newblock Gravitational particle creation and inflation.
\newblock {\em Phys.~Rev.~D}, 35(10):2955--2960, May 1987.

\bibitem{holevo1999evaluating}
A.~S. Holevo and R.~F. Werner.
\newblock Evaluating capacities of bosonic {Gaussian} channels.
\newblock {\em Phys.~Rev.~A}, 63(3):032312, February 2001.

\bibitem{Capozziello_2016}
S.~Capozziello, O.~Luongo, and M.~Paolella.
\newblock Bounding $f({R})$ gravity by particle production rate.
\newblock {\em Int.~J.~Mod.~Phys.~D}, 25(04):1630010, March 2016.

\bibitem{Capozziello:2023vne}
Salvatore Capozziello, Vittorio De~Falco, and Carmen Ferrara.
\newblock {The role of the boundary term in {$f(Q,~B)$} symmetric teleparallel gravity}.
\newblock {\em Eur. Phys. J. C}, 83(10):915, October 2023.

\bibitem{Jim_nez_2018}
J.~B. Jim{\'{e}}nez, L.~Heisenberg, and T.~S. Koivisto.
\newblock Teleparallel palatini theories.
\newblock {\em J.~Cosmol.~Astrop.~P.}, 2018(08):039--039, August 2018.

\bibitem{Bahamonde_2022}
S.~Bahamonde and L.~Järv.
\newblock Coincident gauge for static spherical field configurations in symmetric teleparallel gravity.
\newblock {\em Eur.~Phys.~J.~C}, 82(10):963, October 2022.

\bibitem{Zeldovich:1971mw}
Y.~B. Zeldovich and A.~A. Starobinsky.
\newblock {Particle production and vacuum polarization in an anisotropic gravitational field}.
\newblock {\em Zh. Eksp. Teor. Fiz.}, 61:2161--2175, June 1971.

\bibitem{Anagnostopoulos_2023}
F.~K. Anagnostopoulos, V.~Gakis, E.~N. Saridakis, and S.~Basilakos.
\newblock New models and big bang nucleosynthesis constraints in $f({Q})$ gravity.
\newblock {\em Eur.~Phys.~J.~C}, 83(1), January 2023.

\bibitem{Benetti:2020hxp}
Micol Benetti, Salvatore Capozziello, and Gaetano Lambiase.
\newblock {Updating constraints on {$f(T)$} teleparallel cosmology and the consistency with Big Bang Nucleosynthesis}.
\newblock {\em Mon. Not. Roy. Astron. Soc.}, 500(2):1795--1805, June 2020.

\bibitem{Capozziello:2017bxm}
S.~Capozziello, G.~Lambiase, and E.~N. Saridakis.
\newblock {Constraining {$f(T)$} teleparallel gravity by Big Bang Nucleosynthesis}.
\newblock {\em Eur. Phys. J. C}, 77(9):576, August 2017.

\bibitem{Jain_2007}
P.~Jain, S.~Weinfurtner, M.~Visser, and C.~W. Gardiner.
\newblock Analog model of a {Friedmann}-{Robertson}-{Walker} universe in {Bose}-{Einstein condensates}: Application of the classical field method.
\newblock {\em Phys.~Rev.~A}, 76(3):033616, September 2007.

\bibitem{Barnaby_2009}
N.~Barnaby and Z.~Huang.
\newblock Particle production during inflation: Observational constraints and signatures.
\newblock {\em Phys.~Rev.~D}, 80(12), December 2009.

\bibitem{Kolb_2023}
E.~W. Kolb, S.~Ling, A.~J. Long, and R.~A. Rosen.
\newblock Cosmological gravitational particle production of massive spin-2 particles.
\newblock {\em J.~High Energy Phys.}, 181(5), May 2023.

\bibitem{Cembranos2023}
J.~Cembranos, L.~Garay, Á. Parra-López, and J.~Sánchez~Velázquez.
\newblock Late vacuum choice and slow roll approximation in gravitational particle production during reheating.
\newblock {\em J.~Cosmol.~Astrop.~P.}, 2023(08):060, August 2023.

\bibitem{Li_2019}
L.~Li, T.~Nakama, C.~M. Sou, Y.~Wang, and S.~Zhou.
\newblock Gravitational production of superheavy dark matter and associated cosmological signatures.
\newblock {\em J.~High Energy Phys.}, 2019(67), July 2019.

\bibitem{Li_2020}
L.~Li, S.~Lu, Y.~Wang, and S.~Zhou.
\newblock Cosmological signatures of superheavy dark matter.
\newblock {\em J.~High Energy Phys.}, 2020(231), July 2020.

\bibitem{Safari_2022}
Z.~Safari, K.~Rezazadeh, and B.~Malekolkalami.
\newblock Structure formation in dark matter particle production cosmology.
\newblock {\em Phys.~Dark Universe}, 37:101092, September 2022.

\bibitem{Laurat_2005}
J.~Laurat et~Al.
\newblock Entanglement of two-mode gaussian states: characterization and experimental production and manipulation.
\newblock {\em J.~Opt.~B}, 7(12):S577--S587, November 2005.

\end{thebibliography}
\end{document}